\documentclass[referee,11pt]{sn-jnl}

\usepackage{amssymb}

\usepackage{bm}
\usepackage{booktabs}
\usepackage[table,xcdraw]{xcolor}
\usepackage{amsmath}
\usepackage{graphicx,subcaption,ragged2e}
\usepackage{hyperref}
\usepackage{multirow}%
\usepackage{amssymb,amsfonts}%
\usepackage{amsthm}%
\usepackage{mathrsfs}%
\usepackage[title]{appendix}%
\usepackage{textcomp}%
\usepackage{manyfoot}%
\usepackage{booktabs}%
\usepackage{algorithm}%
\usepackage{algorithmicx}%
\usepackage{algpseudocode}%
\usepackage{listings}%
\usepackage{multicol}

\usepackage{caption}

\newcommand\norm[1]{\lVert#1\rVert}

\newcommand{\Real}{\mathbb{R}}

\newcommand{\sd}{{\sf d}}

\def\mbR{\mathbb{R}}
\def\mR{\mathcal{R}}

\def\mL{\mathcal{L}}
\def\mbN{\mathbb{N}}
\def\mN{\mathcal{N}}
\def\mbZ{\mathbb{Z}}

\def\mbE{\mathbb{E}}
\def\mE{\mathcal{E}}
\def\rw{\rightarrow}
\def\mD{\mathcal{D}}

\def\mT{\mathcal{T}}

\def\bM{\bm{M}}

\newtheorem{Proposition}{\em Proposition}

\newtheorem{Assumption}{\em Assumption}
\newtheorem{Remark}{\em Remark}

\newcommand{\Prob}{\text{Prob}}

\newcommand{\beq}{\begin{equation}}
\newcommand{\eeq}{\end{equation}}
\newcommand{\beqa}{\begin{eqnarray}}
\newcommand{\eeqa}{\end{eqnarray}}
\newcommand{\nn}{\nonumber}

\begin{document}

\title[Maximum likelihood inference for a class of discrete-time Markov-switching time series models with multiple delays]{Maximum likelihood inference for a class of discrete-time Markov-switching time series models with multiple delays}

\author[1]{\fnm{Jos\'e A.}\sur {Mart\'inez-Ordo\~nez}}\email{joseamar@pa.uc3m.es}

\author[1]{\fnm{Javier} \sur{L\'opez-Santiago}}\email{jalopezs@ing.uc3m.es}

\author*[1]{\fnm{Joaqu\'in} \sur{Miguez}}\email{jmiguez@ing.uc3m.es}

\affil[1]{\orgdiv{Departament of Signal Theory and Communications}, \orgname{Universidad Carlos III de Madrid},
            \orgaddress{\country{Spain}}
}
\abstract{Autoregressive Markov switching (ARMS) time series models are used to represent real-world signals whose dynamics may change over time. They have found application in many areas of the natural and social sciences, as well as in engineering. In general, inference in this kind of systems involves two problems: (a) detecting the number of distinct dynamical models that the signal may adopt and (b) estimating any unknown parameters in these models. In this paper, we introduce a class of ARMS time series models that includes many systems resulting from the discretisation of stochastic delay differential equations (DDEs). Remarkably, this class includes cases in which the discretisation time grid is not necessarily aligned with the delays of the DDE, resulting in discrete-time ARMS models with real (non-integer) delays. We describe methods for the maximum likelihood detection of the number of dynamical modes and the estimation of unknown parameters (including the possibly non-integer delays) and illustrate their application with an ARMS model of El Ni\~no--southern oscillation (ENSO) phenomenon.}


\keywords{Markov switch, time series models, delay difference equations, maximum likelihood estimation, expectation maximization algorithm, ENSO}

\maketitle

%
\section{Introduction}
\label{sIntro}

%
\subsection{Background}

Discrete-time autoregressive Markov switching (ARMS) time series models \cite{Franke12} are used to represent real-world signals whose dynamics may change over time. For example, if $\{x_n\}$ is the signal of interest, we may model its evolution as 
\beq
x_n = \phi(x_{0:n-1},u_n,l_n),
\label{eqModel0}
\eeq
where $n \ge 0$ is the current time, $x_{0:n-1}=\{ x_0, x_1, \ldots, x_{n-1} \}$ is the signal history, $\{u_n\}$ is some noise (random) process and $\{l_n\}$ is a Markov chain \cite{Robert04}, i.e., a sequence of discrete random indices that change over time according to a Markov kernel that describes the conditional probabilities $\Prob\left( l_n=i | l_{n-1}=j \right)$ for suitable integers $i$ and $j$. If $l_n \ne l_{n-1}$ then the functions $\phi(\cdot,\cdot,l_n)$ and $\phi(\cdot, \cdot, l_{n-1})$ are different as well, and hence the dynamics of $x_n$ change. ARMS models have found a plethora of applications in statistical signal processing for econometrics \cite{Sims08,Guerin13,Cavicchioli14,Casarin18,Fernandez22}, engineering \cite{Fu10,Fritsche13,Pulford15,Magnant15}, the Earth sciences \cite{Ailliot12,Monbet17,Latimier20}, or complex networks \cite{Lacasa18}, to name a few examples.

Inference for ARMS models involves 
\begin{itemize}
\item[(a)] the detection of the number of possible values that the Markov chain $\{l_n\}$ can take, 
\item[(b)] and the estimation of unknown parameters in each function $\phi(\cdot,\cdot,l_n)$.
\end{itemize}
We may assume, without loss of generality, that $l_n \in \{ 1, \ldots, L \}$. Following \cite{Lacasa18}, in this paper we refer to each dynamical model $\phi(\cdot,\cdot,l)$, $1 \le l \le L$, as a {\em layer} and, hence, task (a) consists in estimating the number of active dynamical layers $L$ from a sequence of data samples $x_{0:n}$. This is a model order selection problem that can be tackled using the Akaike and Bayesian information criteria (AIC and BIC, respectively) \cite{Leroux92, Ryden95, Psaradakis03, Latimier20}, penalised distances \cite{MacKay02}, penalised likelihoods \cite{Cappe05,Monbet17}, the three-pattern method \cite{Psaradakis03} or the Hannan-Quinn criterion (HQC) \cite{Psaradakis03}. Linear ARMS models admit an equivalent representation as autoregressive moving-average (ARMA) systems and, in this case, the number of layers $L$ can also be inferred from the covariance matrix of the ARMA process \cite{Zhang01,Cavicchioli14}.

As for problem (b), maximum likelihood (ML) and maximum a posteriori estimators can be approximated using different forms of the expectation-maximisation (EM) algorithm \cite{Franke12,Ailliot12,Monbet17}, while Markov chain Monte Carlo (MCMC) methods have been applied for Bayesian estimation \cite{Sims08,Magnant15,Zheng17,Casarin18,Lacasa18}. Simpler moment matching techniques can also be applied for parameter estimation in linear ARMS systems \cite{Hocht09}.

See \cite{Franke12,Phoong22} for a survey of recent ARMS models and methods.

\subsection{Contributions}

In this paper we investigate nonlinear ARMS models where the independent noise process $\{u_n\}$ is additive and each dynamical layer depends on a different {\em delay} of the signal. For example, we may rewrite Eq. \eqref{eqModel0} as 
\beq
x_n = \phi(x_{n-1},x_{n-D[l_n]},l_n) + \sigma[l_n]u_n,
\label{eqModel1}
\eeq
where $\{l_n\}$ is a homogeneous Markov chain and each $D[l] > 1$ ($1 \le l \le L$) is a (possibly long) delay, specific to the $l$-th dynamical layer. A detailed description of the relevant family of models is given in Section \ref{sTSModels} below. Our formulation is devised to target time series models that result from the discretisation of delay differential equations \cite{Bellen13,Buckwar00}, which appear often in geophysics \cite{Keane17,Wang18enso}.

The specific contributions of this work can be summarised as follows:
\begin{itemize}
\item We introduce an ARMS time series model that includes systems resulting from the discretisation of stochastic DDEs. In particular, the proposed model includes cases in which the times at which the signal can be observed are not necessarily aligned with the relevant delays (which are often unknown) resulting in discrete-time models with real (non-integer) delays.
\item We extend classical results of \cite{Yao00} on first-order ARMS models to provide sufficient conditions for the new nonlinear ARMS models with multiple delays to be stable.
\item We provide an EM framework for parameter estimation, based on space alternation and a simple stochastic optimisation algorithm, that can be systematically implemented for the models of the proposed class. This scheme can be easily combined with an ML detector of the number $L$ of dynamical layers.
\item We illustrate the application of the proposed model and inference methodology by discretising and then fitting a stochastic DDE which has been proposed as a representation of El Ni\~no--southern oscillation (ENSO) \cite{Wang18enso}. We obtain numerical results for models with up to three dynamical layers and either integer or real delays. In this example non-integer delays appear naturally when the observation times are not aligned with the {\em physical} delays. We validate the model and estimation algorithm using synthetically generated observations and then apply the methodology to real ENSO data.
\end{itemize}

%
\subsection{Notation}

Scalar magnitudes are denoted by regular-face letters, e.g., $x$. Column vectors and matrices are represented by bold-face letters, either lower-case or upper-case, respectively. For example, $\bm{x} = \left( x_1, \ldots, x_m \right)^\top$ is an $m \times 1$ vector (the superscript $^\top$ denotes transposition) and $\bm{X} = \left( \bm{x}_1, \ldots, \bm{x}_d \right)$ is an $m \times d$ matrix, with $\bm{x}_i = \left( x_{1i}, \ldots, x_{mi} \right)^\top$ denoting its $i$-th column. Discrete-time is indicated as a subscript, e.g., $\bm{x}_n$. Dependences on an integer index other than time are represented with the index between square brackets, e.g., $D[l]$ in \eqref{eqModel1} is the delay associated to the $l$-th dynamical layer.

We abide by a simplified notation for probability functions, where $p(x)$ denotes the probability density function (pdf) of the random variable (r.v.) $x$. This notation is argument-wise, hence if we have two r.v.'s $x$ and $y$, then $p(x)$ and $p(y)$ denote the corresponding density functions, possibly different; $p(x,y)$ denotes the joint pdf and $p(x|y)$ is the conditional pdf of $x$ given $y$. The notation for multidimensional r.v.'s, e.g., $\bm{x}$ and $\bm{y}$, is analogous, i.e., $p(\bm{x},\bm{y})$, $p(\bm{x}|\bm{y})$, etc. The probability mass function (pmf) of a discrete r.v. $x$ is denoted $P(x)$ (note the upper case) and we follow the same argument-wise convention as for pdf's.

%
\subsection{Contents}

The rest of the paper is organised as follows. In Section \ref{sTSModels} we introduce the new class of discrete-time, nonlinear ARMS models with multiple delays and provide sufficient conditions that ensure convergence to a limit distribution. An EM framework for inference is described in Section \ref{sEstimation} and computer simulation examples are presented in Section \ref{sExamples}. A case study with real ENSO sea-surface temperature anomalies is presented in Section \ref{sRealData}. Section \ref{sConclusions} is devoted to the conclusions.

%
\section{Time series models}
\label{sTSModels}

%
\subsection{Delayed nonlinear ARMS time series models}

We introduce a nonlinear ARMS time series model with $L$ layers, i.e., $L$ different dynamical modes, each one induced by a different nonlinear map and a different integer delay. An extended model with non-integer delays is described in Section \ref{ssCDMS} below.

Let $\{\bm{x}_n\}$ be a random sequence taking values in $\mbR^d$, and let $\{l_n\}$ be a homogeneous Markov chain, taking values on the finite set $\mL = \{1, \ldots, L\}$, with $L \times L$ transition matrix denoted as $\bm{M}$ and initial pmf $P_0:\mL \mapsto [0,1]$. The entry in the $i$-th row and $j$-th column of $\bm{M}$, denoted $M_{ij}$, is the probability mass $P(l_n=j | l_{n-1}=i)$. The delayed nonlinear ARMS model with $L$ layers, denoted DN-ARMS($L$), is constructed as
\begin{equation}
\bm{x}_n = \phi[l_n](\bm{x}_{n-1}, \bm{x}_{n - D[l_n]}, \bm{\alpha}) + \bm{u}_n[l_n],
\quad n \ge 0,
\label{eq1}
\end{equation}
where $\bm{\alpha} = (\alpha_1, \ldots, \alpha_k )^\top$ is a $k \times 1$ vector of real model parameters, $\Lambda_D = \{ D[1], \ldots, D[L] \}$ is a set of positive integer delays, one per dynamical layer, the functions $\phi[1], \ldots, \phi[L]$ are distinct $\mbR^d \times \mbR^d \times \mbR^k \mapsto \mbR^d$ nonlinear maps, and $\bm{u}_n[1], \ldots, \bm{u}_n[L]$ are independent sequences of $d\times 1$ i.i.d. random vectors with layer-dependent distinct pdf's $p_l(\bm{u})$, $l=1, \ldots, L$. The model description is complete with a prior pdf $p(\bm{x}_{-D^+:-1})$, where $D^+ = \max_{l \in \mL} D[l]$ is the maximum delay and $\bm{x}_{i:j}$ denotes the subsequence $\bm{x}_i, \bm{x}_{i+1}, \ldots, \bm{x}_j$. Note that a distribution for $\bm{x}_{-1}$ is not sufficient to specify the model because of the (possibly long) delays.


%
\subsection{Continuous-delay Markov-switch nonlinear models}
\label{ssCDMS}

Model \eqref{eq1} can be extended to incorporate real positive delays. Non-integer delays arise, for example, from the discretisation of stochastic DDEs when the continuous-time delays are not aligned with the time grid of the observed series. Such scenarios are quite natural. For example, in Section \ref{ssENSO} we look into models of ENSO temperature anomalies. These temperatures are typically collected on a monthly basis; however, there is no physical reason for the delays in the DDE models to be an integer number of months. 

Assume that $D[l] \in (1,+\infty)$ for all $l \in \mL$. Model \eqref{eq1} can be extended to account for possibly non-integer delays if we rewrite it as
\begin{equation}
\bm{x}_n = \phi[l_n](\bm{x}_{n-1}, \tilde{\bm{x}}_{n - D[l_n]}, \bm{\alpha}) + \bm{u}_n[l_n],
\quad n \ge 0,
\label{eq2}
\end{equation}
where $\tilde{\bm{x}}_{n-D[l]}$ is constructed as an interpolation of consecutive elements of the series $\bm{x}_n$. In general, for $\tau \in \mbR^+$, we let
$
\tilde{\bm{x}}_{\tau} = \sum_{m=0}^\infty \kappa(\tau-m) \bm{x}_m,
$
where $\kappa : \mbR \mapsto \mbR$ is an interpolation kernel satisfying that $\tilde{\bm{x}}_\tau = \bm{x}_\tau$ when $\tau$ is an integer. In the computer experiments of Sections \ref{sExamples} and \ref{sRealData} we restrict our attention, for simplicity, to the order 1 interpolation
\beq
\tilde{\bm{x}}_\tau = (\lfloor \tau \rfloor + 1 - \tau) \bm{x}_{\lfloor \tau \rfloor} + (\tau - \lfloor \tau \rfloor) \bm{x}_{\lfloor \tau \rfloor + 1}, 
\label{eqLinInterp}
\eeq
where, for a real number $r\in\mbR$, $\lfloor r \rfloor = \sup\{n \in \mbZ: n<r \}$.  

Let us remark that $\tilde{\bm{x}}_{n-D[l_n]}$ in Eq. \eqref{eq2} is not an observed data point, however it can be deterministically computed from observed data. Also, model \eqref{eq2} reduces to model \eqref{eq1} when $D[1], \ldots, D[L]$ are all integers. We refer to model \eqref{eq2}, with real delays, as cDN-ARMS($L$).

\label{ssDMS}
%
\subsection{Invariant distribution and mean power}
\label{ssErgodicity}

The classical results of \cite{Yao00} on the ergodicity and invariant distributions of first-order ARMS models with additive noise can be extended to the cDN-ARMS($L$) model described in Section \ref{ssCDMS} above. Let us fix that the model parameters $\bm{\alpha}$ and define the $\mbR^d\times\mbR^d\mapsto\mbR^d$ functions
\beq
\phi_{\bm{\alpha}}[l](\bm{x},\bm{x}') := \phi[l](\bm{x},\bm{x}',\bm{\alpha}),
\quad \text{for $l=1, \ldots, L$.}
\nn
\eeq
We make the following assumptions on the cDN-ARMS($L$) model:
\begin{Assumption}
\label{asL}
The functions $\phi_{\bm{\alpha}}[1],\ldots,\phi_{\bm{\alpha}}[L]$ are Lispchitz. In particular, there are finite constants $K_{\bm{\alpha}}[1], \ldots, K_{\bm{\alpha}}[L]$ such that 
$$
\| \phi_{\bm{\alpha}}[l](\bm{x},\bm{x}') - \phi_{\bm{\alpha}}[l](\bm{y},\bm{y}') \| \le K_{\bm{\alpha}}[l] \sqrt{
	\| \bm{x}-\bm{y} \|^2 + \| \bm{x}'-\bm{y}' \|^2
},
$$ 
for $l=1, \ldots, L$, where $\| \cdot \|$ denotes the Euclidean norm. 
\end{Assumption}
\begin{Assumption}
\label{asM}
The Markov chain $\{l_n\}$ with transition matrix $\bm{M}$ is stationary, with invariant distribution $P_\infty(l)$, $l \in \mL$.
\end{Assumption}
\begin{Assumption}
\label{asLM}
The invariant distribution $P_\infty$ and the Lipschitz constants $K_{\bm{\alpha}}[l]$, $l \in \mL$, satisfy the inequality $\sum_{l=1}^L P_\infty(l) \log\left( \sqrt{2}\hat{K}_{\bm{\alpha}}[l] \right) < 1$, where $\hat{K}_{\bm{\alpha}}[l] = 1 \vee K_{\bm{\alpha}}[l]$.
\end{Assumption}

The results stated below follow from Theorems 3 and 4 in \cite{Yao00}. 
\begin{Proposition}
\label{prop1}
Let $\eta_n$ denote the probability law of $\bm{x}_n$. If Assumptions \ref{asL}, \ref{asM} and \ref{asLM} hold, then there exist a unique probability law $\eta$ such that $\lim_{n\rightarrow\infty} \eta_n = \eta$.
\end{Proposition}
\begin{proof}
We convert model \eqref{eq2} with delays for $\bm{x}_n$ in $\mbR^d$ into a first-order model (without delays) in a higher dimensional space. Specifically, let us define
\beq
\bm{y}_n := \left(
	\begin{array}{c}
	\bm{x}_n\\
	\bm{x}_{n-1}\\
	\vdots\\
	\bm{x}_{n-S}\\
	\end{array}
\right), \quad
\bar\phi_{\bm{\alpha}}[l](\bm{y}_{n-1}) := \left(
	\begin{array}{c}
	\phi_{\bm{\alpha}}[l](\bm{x}_{n-1},\bm{x}_{n-D[l_n]})\\
	\bm{x}_{n-1}\\
	\vdots\\
	\bm{x}_{n-S}\\
	\end{array}
\right), 
\label{eqYPhi}
\eeq
and
\beq
\bm{v}_n[l_n] = \left(
	\begin{array}{c}
	\bm{u}_n[l_n]\\
	\bm{0}\\
	\vdots\\
	\bm{0}\\
	\end{array}
\right),
\nn
\eeq
where $\bm{y}_n, \bm{v}_n \in \mbR^{d(S+1)}$, $S=\inf \left\{ n\in \mbN: n\ge \max_{1\le l \le L} D[l] \right\}$ and $\bar{\phi}_{\bm{\alpha}}:\mbR^{d(S+1)}\mapsto\mbR^{d(S+1)}$. Model \eqref{eq2} can now be equivalently rewritten as
\beq
\bm{y}_n = \bar{\phi}_{\bm{\alpha}}[l_n](\bm{y}_{n-1}) + \bm{v}_n[l_n],
\label{eqHDim}
\eeq
where the functions $\bar{\phi}_{\bm{\alpha}}[l]$ are Lipschitz. Indeed, from Eq \eqref{eqYPhi} and Assumption \ref{asL} we readily see that
\beqa
\left\|
	\bar{\phi}_{\bm{\alpha}}[l](\bm{y}_{n-1}) - \bar{\phi}_{\bm{\alpha}}[l](\bm{y}_{n-1}')
\right\|^2 
&\le& K_{\bm{\alpha}}[l]^2 \left(
	\| \bm{x}_n - \bm{x}_n' \|^2 + \| \bm{x}_{n-D[l]} - \bm{x}_{n-D[l]}' \|^2
\right) \nn\\
&& + \sum_{j=1}^S \| \bm{x}_{n-j} - \bm{x}_{n-j}' \|^2.
\label{eqP0-1}
\eeqa
If we let $\hat{K}_{\bm{\alpha}}[l] = \left( K_{\bm{\alpha}}[l] \vee 1 \right)$ then it is apparent that \eqref{eqP0-1} implies
\beq
\left\|
	\bar{\phi}_{\bm{\alpha}}[l](\bm{y}_{n-1}) - \bar{\phi}_{\bm{\alpha}}[l](\bm{y}_{n-1}')
\right\|^2 \le 2\hat{K}_{\bm{\alpha}}[l]^2 \sum_{j=0}^S \| \bm{x}_{n-j} - \bm{x}_{n-j}' \|^2
\nn
\eeq
and, therefore,
$
\left\|
	\bar{\phi}_{\bm{\alpha}}[l](\bm{y}_{n-1}) - \bar{\phi}_{\bm{\alpha}}[l](\bm{y}_{n-1}')
\right\| \le \sqrt{2}\hat{K}_{\bm{\alpha}}[l] \| \bm{y} - \bm{y}' \|,
$
i.e., $\bar{\phi}_{\bm{\alpha}}[l]$ is Lipschitz with constant $\sqrt{2}\hat{K}_{\bm{\alpha}}[l]$. Combining this result with Assumptions \ref{asM} and \ref{asLM} we can apply Theorem 3 in \cite{Yao00}, which implies the statement of Proposition \ref{prop1}. 
\end{proof}

Let us additionally introduce the matrix 
$$
\bm{M}_s := \begin{pmatrix}
    (\sqrt{2} K_{\bm{\alpha}}[1])^s M_{1,1} & \cdots & (\sqrt{2} K_{\bm{\alpha}}[l])^s M_{1,l} \\
    \vdots & \ddots & \vdots \\
    (\sqrt{2} K_{\bm{\alpha}}[1])^s M_{l,1} & \cdots & (\sqrt{2} K_{\bm{\alpha}}[l])^s M_{l,l} \\
    \end{pmatrix},
$$
where $s \ge 1$, and let $\rho(\bm{M}_s)$ denote the spectral radius of $\bm{M}_s$. Then, we can characterise the moments of order $s$ of the sequence $\bm{x}_n$.

\begin{Proposition}
\label{prop2}
If Assumptions \ref{asL} and \ref{asM} hold, and $\rho(\bm{M}_s)<1$, then the unique law $\eta=\lim_{n\rightarrow\infty} \eta_n$ has finite moments of order $s$. In particular, if $\rho(\bm{M}_2)<1$ then $\lim_{n\rw\infty}\mbE\left[ \| \bm{x}_n \|^2 \right] < \infty$.
\end{Proposition} 
\noindent \textit{\textbf{Proof:}} This result follows from Theorem 4 in \cite{Yao00} by the same argument as in the proof of Proposition \ref{prop1}. \qed

%
\section{Model inference}
\label{sEstimation}

We introduce a space-alternating (SA) EM algorithm for iterative ML parameter estimation in the general cDN-ARMS($L$) model described in Section \ref{ssCDMS}. First, we obtain the likelihood for the proposed class of models, recall the standard EM method and explain why it is not tractable. We then describe the SA-EM scheme and conclude this section with a succinct discussion on the ML detection of the number $L$ of dynamical layers.

%
\subsection{Likelihood function}
\label{ssLikelihood}

Let $\bm{x}_{0:T}=\{ \bm{x}_0, \ldots, \bm{x}_T \}$ be a sequence of observations (and assume that $\bm{x}_n$ is given for all $n<0$). The set of model parameters to be estimated is $\Lambda = \Lambda_M \cup \Lambda_D \cup \Lambda_\alpha$, where
$$
\Lambda_M = \{ M_{1,1}, \ldots, M_{L,L} \},~~
\Lambda_D = \{ D[1], \ldots, D[L] \}, \text{~and~}
\Lambda_\alpha = \{ \alpha_1, \ldots, \alpha_k \}
$$
are the set of entries of the $L\times L$ transition matrix $\bm{M}$, the set of (possibly real) delays and the set of entries of the $k\times 1$ vector $\bm{\alpha}$, respectively. 

We denote the likelihood of the parameter set $\Lambda$ given the observed sequence $\bm{x}_{0:T}$ as $p(\bm{x}_{0:T} | \Lambda)$. In order to obtain an explicit expression for the likelihood we write it in terms of the joint distribution of $\bm{x}_{0:T}$ and the Markov sequence of layers $l_{0:T} = \{ l_0, \ldots, l_T \}$, namely,
\beq
p(\bm{x}_{1:T}|\Lambda)  =  \sum_{l_{0:T} \in \mL^{T+1}} p(\bm{x}_{0:T},l_{0:T}|\Lambda).
\label{eqMarginal}
\eeq
For any $0 < n \le T$, we can use Bayes' theorem to obtain a recursive decomposition of the joint distribution, 
\beq
p(\bm{x}_{0:n}, l_{0:n}|\Lambda) = p(\bm{x}_n| \bm{x}_{0:n-1},l_n,\Lambda) M_{l_{n-1},l_n} p(\bm{x}_{0:n-1}, l_{1:n-1}|\Lambda)
\label{eqRec}
\eeq
where we have used the Markov property and the fact that $l_n$ is conditionally independent of $\bm{x}_{0:n-1}$ to show that $p(l_n|l_{1:n-1},\bm{x}_{1:n-1},\Lambda) = M_{l_{n-1},l_n}$ and $p(\bm{x}_n| \bm{x}_{0:n-1},l_{0:n},\Lambda) = p(\bm{x}_n| \bm{x}_{0:n-1},l_n,\Lambda)$. If we (repeatedly) substitute the recursive relationship \eqref{eqRec} into Eq. \eqref{eqMarginal} we readily obtain
\beq
p(\bm{x}_{1:T}|\Lambda)  =  \sum_{l_{0:T} \in \mL^{T+1}} \prod_{n=1}^T p(\bm{x}_n| \bm{x}_{0:n-1},l_n,\Lambda) M_{l_{n-1},l_n} p(\bm{x}_0|l_0,\Lambda) P_0(l_0),
\nn
\eeq
where all factors can be computed. In particular, note that $p(\bm{x}_0|l_0,\Lambda)$ is tractable because we have assumed that $\bm{x}_{-1}, \bm{x}_{-2}, \ldots$ are known.

The ML estimator of the parameters is the solution of the problem $\Lambda_{ML} \in \arg\max_{\Lambda} p(\bm{x}_{0:T}|\Lambda)$. However, even when the $\phi[l]$'s are linear, it is not possible to compute $\Lambda_{ML}$ exactly and we need to resort to numerical approximations \cite{Franke12}.

%
\subsection{Expectation-maximisation algorithm}
\label{ssEM}


Let $x$ and $y$ be r.v.'s, let $\theta$ be some parameter and let $f(x)$ be some transformation of $x$. We write $\mbE_{x|y,\theta}[ f(x) ]$ to denote the expected value of $f(x)$ w.r.t. the distribution with pdf $p(x|y,\theta)$, i.e., 
\beq
\mbE_{x|y,\theta}[ f(x) ] = \int f(x) p(x|y,\theta) \sd x.
\nn
\eeq

A standard EM algorithm \cite{Dempster77,McLachlan07} for the iterative ML estimation of $\Lambda$ from the data sequence $\bm{x}_{0:T}$ can be outlined as follows.
\begin{enumerate}
\item \textit{Initialisation:} choose an initial (arbitrary) estimate $\hat{\Lambda}_0$.
\item \textit{Expectation step:} given an estimate $\hat{\Lambda}_i$, obtain the expectation 
\beq
\mE_i(\Lambda) = \mbE_{l_{0:T}|\bm{x}_{0:T}, \hat{\Lambda}_i}\left[  
    \log\left(
        p(\bm{x}_{0:T},l_{0:T}|\Lambda)
    \right)
\right].
\label{eqStandardE}
\eeq
\item \textit{Maximisation step:} obtain a new estimate,
\beq
\hat{\Lambda}_{i+1} \in \arg\max_{\Lambda} \mE_i(\Lambda).
\label{eqStandardM}
\eeq
\end{enumerate}
Using standard terminology, $\bm{x}_{0:n}$ is the observed data, $l_{0:T}$ is the latent data and $\{ \bm{x}_{0:T},l_{0:T} \}$ is the complete data set. The basic guarantee provided by the EM algorithm is that the estimates $\hat{\Lambda}_0, \hat{\Lambda}_1, \ldots$ have non-decreasing likelihoods, i.e., for every $i \ge 0$, $p(\bm{x}_{0:T}|\hat{\Lambda}_{i+1}) \ge p(\bm{x}_{0:T}|\hat{\Lambda}_i)$ \cite{McLachlan07}.

If we substitute the recursive decomposition \eqref{eqRec} into the expectation of \eqref{eqStandardE} we arrive at
\beq
\mE_i(\Lambda) = \mE_i^0(\Lambda_D \cup \Lambda_\alpha) + \mE_i^1(\Lambda_M) + \log(P_0(l_0)),
\label{eqE}
\eeq
where
$
\mE_i^0(\Lambda_D \cup \Lambda_\alpha) = \sum_{n=0}^{T} \mbE_{l_n|\bm{x}_{0:T},\hat{\Lambda}_i}\left[
    \log(p(\bm{x}_n| \bm{x}_{0:n-1}, l_n, \Lambda_D \cup \Lambda_\alpha))
\right]
$
and
$
\mE_i^1(\Lambda_M) = \sum_{n = 1}^T \mbE_{l_{n-1:n}|\bm{x}_{1:T},\hat{\Lambda}_i}\left[
    \log(M_{l_{n-1},l_n})
\right]
$.
Now, if we write the posterior expectations $\mbE_{l_n|\bm{x}_{0:T},\hat{\Lambda}_i}[\cdot]$ and $\mbE_{l_{n-1:n}|\bm{x}_{1:T},\hat{\Lambda}_i}[\cdot]$ explicitly we obtain
\beqa
\hspace{-0.6cm}
\mE_i^0(\Lambda_D \cup \Lambda_\alpha) &=& \sum_{n=0}^{T} \sum_{l_n \in \mL} 
    \log(p(\bm{x}_n| \bm{x}_{0:n-1}, l_n, \Lambda_D \cup \Lambda_\alpha))
    P(l_n|\bm{x}_{0:n},\hat{\Lambda}_i), \label{eqEDa} \\
\mE_i^1(\Lambda_M) &=& \sum_{n = 1}^T \sum_{l_{n-1} \in \mL} \sum_{l_n\in\mL} 
    \log(M_{l_{n-1},l_n})
    P(l_{n-1},l_n|\bm{x}_{0:n},\hat{\Lambda}_i),
\label{eqEM}
\eeqa
where $P(l_n|\bm{x}_{0:n},\hat{\Lambda}_i)$ and $P(l_{n-1},l_n|\bm{x}_{0:n},\hat{\Lambda}_i)$ are the posterior pmf's of the random indices $l_n$ and $l_{n-1:n}$, respectively, conditional on the observations $\bm{x}_{0:n}$ and the $i$-th parameter estimators $\hat{\Lambda}_i$.

The posterior pmf's $P(l_n|\bm{x}_{0:n},\hat{\Lambda}_i)$ and $P(l_{n-1},l_n|\bm{x}_{0:n},\hat{\Lambda}_i)$ can be computed exactly, for every $n=0, \ldots, T$, by running a forward-backward algorithm \cite{Cappe05,Franke12}. Given these probabilities, from the sequence of Eqs. \eqref{eqStandardM}, \eqref{eqE}, \eqref{eqEDa} and \eqref{eqEM} it is apparent that we can update\footnote{We denote $\hat{\Lambda}_{i+1} = \hat{\Lambda}_{M,i+1} \cup \hat{\Lambda}_{D,i+1} \cup \hat{\Lambda}_{\alpha,i+1}$.}
\beq
\hat{\Lambda}_{M,i+1} = \arg\max_{\Lambda^M} \mE_i^1(\Lambda_M).
\nn
\eeq
Unfortunately, the problem 
$
\hat{\Lambda}_{D,i+1} \cup \hat{\Lambda}_{\alpha,i+1} \in \arg\max_{\Lambda_D \cup \Lambda_\alpha} \mE_i^0(\Lambda_D \cup \Lambda_\alpha)
$
is analytically intractable in general and the (typically) large number of parameters in $\Lambda_D \cup \Lambda_\alpha$ makes numerical approximations hard in practice. As a workaround, we propose a SA-EM algorithm \cite{Fessler94} that can be used systematically for the class of models described by \eqref{eq2}. 


%
\subsection{Space-alternating expectation-maximisation algorithm}
\label{ssSAEM}

Let us re-partition the parameter set as $\Lambda = \Lambda_M \cup \Lambda_* \cup \Lambda_c$, where $\Lambda_M$, $\Lambda_*$ and $\Lambda_c$ are disjoint sets, and
\begin{itemize}
\item $\Lambda_M$ contains the $L \times L$ entries of the transition matrix $\bm{M}$, as before;
\item $\Lambda_* \subseteq \Lambda_D \cup \Lambda_\alpha$ contains the delays $D[l]$ and possibly some entries of  $\bm{\alpha}$;
\item and $\Lambda_c = \left( \Lambda_D \cup \Lambda_\alpha \right) \backslash \Lambda_*$ contains the remaining parameters.
\end{itemize}
The partition of $\Lambda_D \cup \Lambda_\alpha = \Lambda_* \cup \Lambda_c$ is not arbitrary. To be specific, $\Lambda_*$ is a subset of $\Lambda_D \cup \Lambda_\alpha$ such that the problem
\beq
\hat{\Lambda}_{*,i+1} = \arg\max_{\Lambda_* \cup \hat{\Lambda}_{c,i}} \mE_i^0(\Lambda_* \cup \hat{\Lambda}_{c,i})
\nn
\eeq
can be solved exactly (note that $\Lambda_* \cup \Lambda_c = \Lambda_D \cup \Lambda_\alpha$, hence $\mE_i^0(\Lambda_* \cup \hat{\Lambda}_{c,i})$ is well defined).

We can now summarise the proposed SA-EM algorithm as follows.
\begin{enumerate}
\item \textit{Initialisation:} choose an initial (arbitrary) estimate $\hat{\Lambda}_0 = \hat{\Lambda}_{M,0} \cup \hat{\Lambda}_{*,0} \cup \hat{\Lambda}_{c,0} = \hat{\Lambda}_{M,0} \cup \hat{\Lambda}_{D,0} \cup \hat{\Lambda}_{\alpha,0}$.

\item \textit{Expectation step:} given an estimate $\hat{\Lambda}_i$, run a forward-backward algorithm to compute the pmf's $P(l_n|\bm{x}_{0:n},\hat{\Lambda}_i)$ for all $l_n \in \mL$ and $P(l_{n-1:n}|\bm{x}_{0:n},\hat{\Lambda}_i)$ for all $l_{n-1:n} \in \mL^2$. Then construct the expectations $\mE_i^1(\Lambda_M)$ and $\mE_i^0(\Lambda_*\cup\Lambda_c)$ in Eqs. \eqref{eqEDa} and \eqref{eqEM}, respectively.

\item \textit{Maximisation step:} compute
\beq
\hat{\Lambda}_{M,i+1} = \arg\max_{\Lambda_M} \mE_i^1(\Lambda_M) 
\nn
\eeq
and denote $\Lambda_c = \{ \lambda^c_1, \ldots, \lambda^c_q \}$ and $\hat{\Lambda}_{c,i} = \{ \lambda^c_{1,i}, \ldots, \lambda^c_{q,i} \}$, where $q \le L+k$ is the number of parameters in $\Lambda_c$. \\For $j = 1, \ldots, q$, compute
\beq
\hat{\lambda}^c_{j,i+1} \text{~~such that~~} \max_{\Lambda_*} \mE_i^0(\Lambda_* \cup \hat{\Lambda}_{c,i+1}^{j+}) \ge \max_{\Lambda_*} \mE_i^0(\Lambda_* \cup \hat{\Lambda}_{c,i+1}^{j-})
\label{eqQ0}
\eeq
where 
\beqa
\hat{\Lambda}_{c,i+1}^{j+} &:=& \left\{ \hat\lambda^c_{1,i+1}, \ldots, \hat\lambda^c_{j-1,i+1}, \hat\lambda^c_{j,i+1}, \hat\lambda^c_{j+1,i}, \ldots, \hat\lambda^c_{q,i} \right\}, \text{~~and~~} \nn\\
\hat{\Lambda}_{c,i+1}^{j-} &:=& \left\{ \hat\lambda^c_{1,i+1}, \ldots, \hat\lambda^c_{j-1,i+1}, \hat\lambda^c_{j,i}, \hat\lambda^c_{j+1,i}, \ldots, \hat\lambda^c_{q,i} \right\},
\nn
\eeqa
to obtain $\hat{\Lambda}_{c,i+1} = \left\{ \hat\lambda^c_{1,i+1}, \ldots, \hat\lambda^c_{q,i+1} \right\}$. Finally, let 
\beq
\hat{\Lambda}_{*,i+1} = \arg\max_{\Lambda_*} \mE_i^0\left(\Lambda_* \cup \hat{\Lambda}_{c,i+1} \right)
\nn
\eeq
and 
$\hat\Lambda_{i+1} = \hat\Lambda_{M,i}\cup\hat\Lambda_{*,i+1}\cup\hat\Lambda_{c,i+1}$.
\end{enumerate}

Intuitively, at the $(i+1)$-th iteration of the algorithm we update the parameters in $\Lambda_M$ exactly, then we update the parameters $\lambda_1^c, \ldots, \lambda^c_q \in \Lambda_c$ one at a time in \eqref{eqQ0}, and finally we update the parameters in $\Lambda_*$ exactly. The 1-dimensional updates in \eqref{eqQ0} can be numerically carried out if various ways. We suggest the application of the accelerated random search (ARS) method of \cite{Appel03} (see also \cite{Marinho07} and \ref{apARS}), which is straightforward to apply and has performed well in our computer experiments (presented in Section \ref{sExamples}).

The SA-EM algorithm can be run for a fixed, prescribed number of iterations or stopped when some criterion is fulfilled, e.g., that the difference between successive estimates is smaller than a given threshold. 


\begin{Remark}
The SA-EM algorithm is designed in the vein of the space-alternating generalised EM methods of \cite{Fessler94}. However, the scheme introduced in this paper does not rigorously below to the class of algorithms described in \cite{Fessler94}. Nevertheless, it is not hard to prove that the estimates $\hat\Lambda_i$ generated by the SA-EM scheme have non-decreasing likelihoods, i.e., $p(\bm{x}_{0:n}|\hat\Lambda_{i+1}) \ge p(\bm{x}_{0:n}|\hat\Lambda_i)$ for every $i \ge 1$. this is the same property enjoyed by the standard EM method and the generalised algorithms of \cite{Fessler94}.
\end{Remark}

%
\subsection{Estimation of the number of layers $L$}
\label{ssEstimateL}

When the number of dynamical layers, $L$, in the cDN-ARMS($L$) model is unknown we can use the proposed SA-EM algorithm to estimate it. In particular, assume that $L \in A:=\{a^-, \ldots, a^+\}$. We can run $I$ iterations of the the SA-EM algorithm for each $L\in A$ to obtain approximate likelihoods
\beq
p(\bm{x}_{0:T}|L) \approx p(\bm{x}_{0:T}|\hat\Lambda_I) =: \ell_T(L), \quad L=a^-, \ldots, a^+,
\nn
\eeq 
where $\hat{\Lambda}_I$ is the set of parameter estimates after $I$ iterations (note that the number of elements in this set increases with $L$). The likelihood $p(\bm{x}_{0:T}|\hat\Lambda_I)$ above can be computed exactly as a by-product of the forward-backward algorithm run in the maximisation step, with a computational cost $\mathcal{O}(T)$ (see \cite{Franke12} for details). 

Choosing the value of $L\in\{a^-,\ldots,a^+\}$ that maximises $\ell_T(L)$ typically leads to overestimation due to over-fitting. Following \cite{Psaradakis03}, we adopt a penalised likelihood estimator of the number of layers. In particular, we have run computer experiments with a penalisation of the likelihood of the form $e^{-C_T |\Lambda|_L}$, where $C_T = \frac{1}{2}\log(T)$ and $|\Lambda|_L$ denotes the number of parameters in the set $\Lambda$ when the model has $L$ layers. This yields the penalised likelihood
$
\tilde \ell_T(L) := e^{-C_T |\Lambda|_L } p(\bm{x}_{0:T}|\hat\Lambda_I)
$
and the penalised estimator
\beq
\hat L = \arg\max_{L\in A} \left\{\log \tilde\ell_T(L)\right\} 
= \arg\max_{L\in A} \left\{ \log p(\bm{x}_{0:T}|\hat\Lambda_I) - C_T |\Lambda|_L \right\}
\nn
\eeq

\section{Computer simulations}
\label{sExamples}


\subsection{El Ni\~no--southern oscillation model}
\label{ssENSO}

El Niño--southern oscillation (ENSO) is a recurring event belonging to a class of climatic phenomena called atmospheric oscillations. It originates from variations in wind intensities in the general atmospheric circulation. These variations cause the oscillation of the thermocline in the Pacific Ocean which, in turn, causes alternating high and low sea surface temperatures (SSTs) on both sides of the ocean. In particular, the El Niño phenomenon corresponds to an increase in SST in the eastern Pacific, which is associated with a strong increase in rainfall intensity and duration in Central and South America. The anomalies in the trade wind and the SST at the Pacific Ocean’s equator have been historically modelled as the solution of a DDE \cite{SS88,PMP97,Wang18enso}. 
Several conceptual equations have also been proposed in the literature \cite{Ghil08enso,JLTZ07} that incorporate stochastic terms or display chaotic dynamics. 

For the computer experiments in this section we consider a nonlinear DDE based on the model of Ghile et al. \cite{Ghil08enso}, where we include a diffusion term to obtain the stochastic DDE in It\^o form
\beq
\sd \mT = \left[
    b \cos({ 2\pi \omega}t) - a \tanh(\kappa \mT(t-\tau)
\right] \sd t + \sigma \sd W,
\label{eq:ghil08}
\eeq
where $t$ denotes continuous time (the time unit is 1 year), $\mT(t)$ is the SST anomaly, $\tau \in \Real^+$ is a time delay, and $a$, $b$, $\kappa$ and $\omega$ are constants, $W(t)$ is a standard Wiener process and $\sigma>0$ is a diffusion factor that determines the intensity of the stochastic perturbation. 
%
%
%

Equation \eqref{eq:ghil08} can be integrated numerically using different schemes \cite{Buckwar00}. For simplicity, we apply a weak-order 1.0 Euler-Maruyama scheme with constant time-step $h>0$, which yields
\beq
\mT_n = \mT_{n-1} + \left[ 
    b \cos\left(
        2\pi \omega h(n-1)
    \right) - a \tanh\left(
        \kappa \mT_{n-1-D}) 
    \right)
\right] h + \sqrt{h} \sigma u_n,
\label{eq:Euler}
\eeq
where $\mT_n \approx \mT(nh)$ is an approximation of the temperature process at $t=nh$, $D$ is a discrete delay computed as $D=\frac{\tau}{h}$ (we assume that $\tau$ can be expressed as an integer multiple of $h$) and $u_n$ is an i.i.d. sequence of $\mN(0,1)$ r.v.'s. 

Model \eqref{eq:ghil08} and its discretised version \eqref{eq:Euler} can be shown to yield sequences of temperatures which are (qualitatively 
 and quantitatively) similar to the measurements of SST anomalies in the South Pacific ocean. However, these models are not accurate enough for reliable forecasting and, in particular, it has not been possible to use them to predict the large ``spikes'' in SST that characterise the El Ni\~no phenomenon. In an attempt at extending the applicability of the model, we propose to construct multi-layer cDN-ARNMS($L$) models based on \eqref{eq:Euler} that extend the flexibility of the original equation. 

%
\subsection{DN-ARMS($L$) model with integer delays}
\label{ssENSO_integer_delays}

Let us construct a DN-ARMS($L$) model of the form in Eq. \eqref{eq1} for SST anomaly time series. Publicly available SST data for ENSO is given in the form of {\em monthly} averaged SSTs and temperature anomalies. For this reason, we adopt $h=\frac{1}{12}$ as a time step in the Euler scheme \eqref{eq:Euler}. This time step is known and common to all $L$ layers of the DN-ARMS($L$) model. Additionally, we define:
\begin{itemize}
\item A Markov chain $\{l_n\}$ taking values in $\mL=\{1, \ldots, L\}$ with an $L \times L$ transition matrix $\bm{M}$.
\item A set of delays $\tau[1], \ldots, \tau[L]$, one per layer. We assume that these delays are integer multiples of $h=\frac{1}{12}$, hence they turn into integer delays $D[1]=\frac{\tau[1]}{h}, \ldots, D[L]=\frac{\tau[L]}{h} \in \mbN$.
\item Parameters $\{ a[l], b[l], \kappa[l], \omega[l], \sigma[l] \}$ for each $l \in \mL$; hence, the parameter vector
$
\bm{\alpha} = \left( a[1], b[1], \ldots, a[L], b[L], \kappa[L], \omega[L], \sigma[L] \right)^\top
$
is $5L \times 1$ (i.e., $k=5L$ in the general model \eqref{eq1}).
\end{itemize}
The assumption $D[l]=\frac{\tau[l]}{h} \in \mbN$ for every $l$ may be a mild one when $h$ can be chosen to be sufficiently small, however it is hardly realistic for a time step of one month (and we drop it in Section \ref{ssENSO_real_delays} below).

The resulting DN-ARMS($L$) model of the ENSO time series can be compactly written as
\beqa
l_n &\sim& P(l_n|l_{n-1}) = M_{l_{n-1},l_n}, \nn\\
x_n &=& x_{n-1} + h\left[ 
    b[l_n] \cos\left(
        2\pi \omega[l] h(n-1)
    \right) - a[l_n] \tanh\left(
        \kappa[l_n] x_{n-D[l_n]}) 
    \right)
\right] \nn \\
&& + \sqrt{h} \sigma[l_n] u_n, \quad n \ge 0,
\label{eq:ENSO-1}
\eeqa
where $x_n$ is the SST anomaly at time $t=nh$ and $\{u_n\}$ is an i.i.d. $\mN(0,1)$ sequence. We assume $x_n \sim \mN(0,1) $ for all $n \le 0$.

\subsubsection{Estimation of the number of layers $L$} 
\label{ssNL}
In the first set of computer experiments we assess the estimation of the number of layers $L$ using the penalised maximum likelihood method described in Section \ref{ssEstimateL}. We simulate two data sets $x_{0:T}^{(2)}$ and $x_{0:T}^{(3)}$ with {\em true} values, $L_o=2$ and $L_o=3$, respectively, and $T=1,000$. Then, for each $L \in \{2,3,4\}$ we run the SA-EM algorithm of Section. Since EM algorithms converge only locally \cite{McLachlan00}, for each $L_o$ and each $L$ we run the SA-EM scheme $R=50$ times, with the same data set but different, independently generated initial parameter estimates $\hat{\Lambda}^{(r)}_{0}, r = 1,\ldots,50$. Recall that the SA-EM algorithm yields the likelihood as a byproduct (see Section \ref{ssEstimateL}). We denote the likelihood of the model with $L$ layers in the $r$-th simulation as $\ell_{T}^{(r)}(L)$ and then assign the ML over the $R$ independent simulations, i.e., $\ell_T(L) = \max_{1\le r \le R} \ell_{T}^{(r)}(L)$. 

The true transition matrices are
$$
\bM = \left[
	\begin{array}{cc}
	0.6 &0.4\\
	0.3 &0.7\\
	\end{array}
\right]
\quad \text{and} \quad
\bM = \left[
	\begin{array}{ccc}
	0.5 & 0.3 & 0.2\\
	0.2 & 0.3 & 0.5\\
        0.2 & 0.6 & 0.2
	\end{array}
\right]
$$
for $L_o=2$ and $L_o=3$, respectively. The remaining parameters are
$$
\begin{array}{lll}
a[1:2] = (10,1)^\top, &\kappa[1:2] = (3,1)^\top, & b[1:2] =(10,1)^\top, \\
w[1:2] = (1/12,1/3)^\top, &\sigma[1:2] = (0.3,0.1)^\top, & D[1:2] = (5,15)^\top,
\end{array}
$$
for $L_o=2$ and, for $L_o=3$,
$$
\begin{array}{lll}
a[1:3] = (10,1,2)^\top, &\kappa[1:3] = (3,2,1)^\top, & b[1:3] =(5,1,3)^\top, \\
w[1:3] = (1/12,1/3,1/5)^\top, &\sigma[1:3] = (0.4,0.2,0.1)^\top, & D[1:3] = (5,10,18)^\top.
\end{array}
$$

\begin{table}[!ht]
    \begin{multicols}{2}
    \centerline{
        \begin{tabular}{|l||l|l|}
        \hline
            ~ &$\ell_T(L)$ &$\tilde\ell_T(L)$\\ \hline\hline
            $L=2$ & 1071.0 & $\bm{1015.7}$ \\ \hline
            $L=3$ & 1086.0 & 992.8 \\ \hline
            $L=4$ & $\bm{1106.7}$ & 968.5 \\ \hline
        \end{tabular}
    }
    \centerline{
        \begin{tabular}{|l||l|l|}
        \hline
            ~ &$\ell_T(L)$ &$\tilde\ell_T(L)$\\ \hline\hline
            $L=2$ & -160.2 & -215.5 \\ \hline
            $L=3$ & 542.2 & $\bm{448.9}$ \\ \hline
            $L=4$ & $\bm{572.0}$ & 433.8 \\ \hline
        \end{tabular}
    }
    \end{multicols}
    \caption{Log-likelihoods and penalised log-likelihoods when the true model has $L_o = 2$ layers (left) and $L_o=3$ layers (right).}
    \label{Ta:Like_2}
\end{table}

Table \ref{Ta:Like_2} shows the log-likelihoods and the penalised log-likelihoods obtained for the models with $L = 2,3,4,$ when the true model generating the data has $L_o=2$ layers (left) and $L_o=3$ layers (right). It is seen that higher likelihoods are obtained as $L$ is increased, even when $L>L_o$. This is due to over-fitting of the model parameters. When a simple penalisation is included, the correct number of dynamical layers is detected in both experiments.

\subsubsection{Parameter estimation} \label{sssParamEst}

In this section we study the accuracy of the SA-EM parameter estimation algorithm with $L_o=2$ layers described in Section \ref{ssNL} and aim at estimating the transition matrix $\bM$ as well as the integer delays $D[1:2] = \left( D[1], D[2] \right)^\top$ and the model parameters $\bm{\alpha} = \left( a[1:2],b[1:2],\kappa[1:2], w[1:2], \sigma[1:2] \right)^\top$. We assess the accuracy of the estimators of real parameters in terms of normalised errors. Assume we run $R$ independent simulations, all with the same true parameters (but independently generated observations $x_{0:T}$). Then, the normalised estimation errors for the transition matrix $\bM$ are
\begin{equation}
    e_{\bm{M}}(r) := \norm{\hat{\bM}^{(r)}-{\bM}}_{F} / \norm{\bM}_{F},
    \quad r=1, \ldots, R,
    \nn
\end{equation}
where $\norm{\cdot}_{F}$ denotes the Frobenius norm and $\hat{\bM}^{(r)}$ represents the estimate of $\bM$ in the $r$-th independent simulation. For the parameters in $\bm{\alpha}$, the normalised errors are computed as 
\begin{equation}
    e_{\alpha_i}(r) = |\hat{\alpha}_i^{(r)}-\alpha_i| / |\alpha_i|
    \quad r=1, \ldots, R,
    \nn
\end{equation}
where $\alpha_i$ denotes the $i$-th entry of vector $\bm{\alpha}$ and $\hat{\alpha}_i^{(r)}$ is its estimate in the $r$-th independent simulation.

Since the delays $D[1:2]$ are integers, calculating a normalised Euclidean norm does not provide a meaningful characterisation of performance. Instead, we assess the estimation algorithm by computing the frequency of correct detections, i.e., if $\hat D[l]^{(r)}$ is the estimate of the delay $D[l]$ in the $r$-th independent simulation, for $r=1, \ldots, R$, then the frequency of (correct) detections is $F_D := \sum_{r=1}^R \delta\left[ \sum_{l=1}^{L} |\hat D[l]^{(r)} - D[l]| \right]$, where $\delta[\cdot]$ is the Kronecker delta function.

\begin{figure}[htb]
\captionsetup[subfigure]{justification=Centering}
    \begin{subfigure}[t]{0.42\textwidth}	
    		\includegraphics[width=\textwidth]{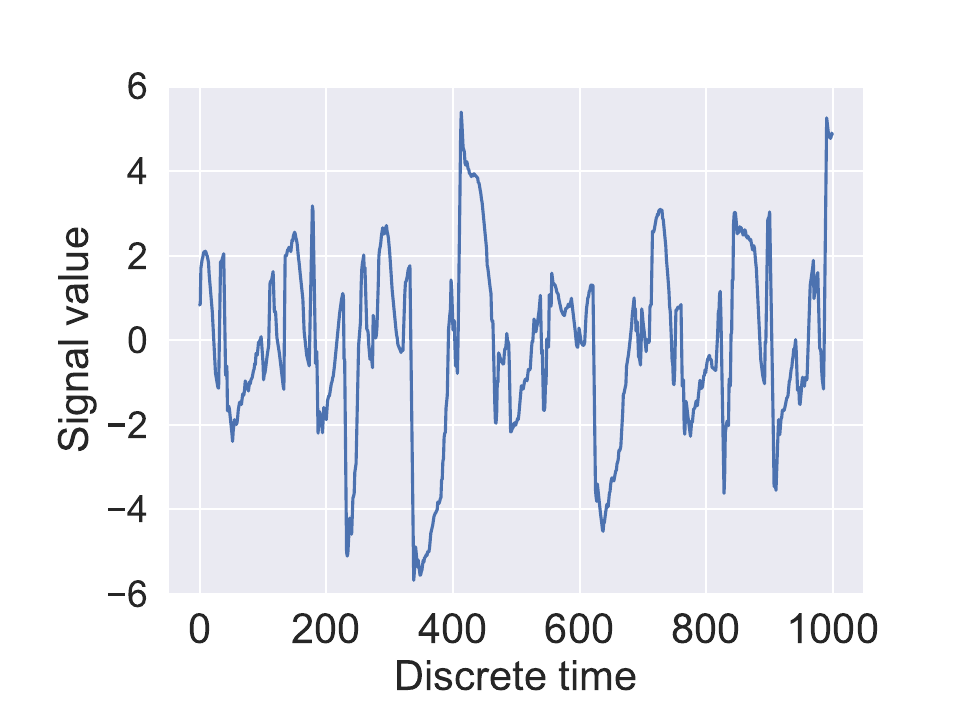}
    		\caption{Sample signal.}
                \label{sub:signal}
    \end{subfigure}\hspace{\fill}
    \begin{subfigure}[t]{0.42\textwidth}
    		\includegraphics[width=\linewidth]{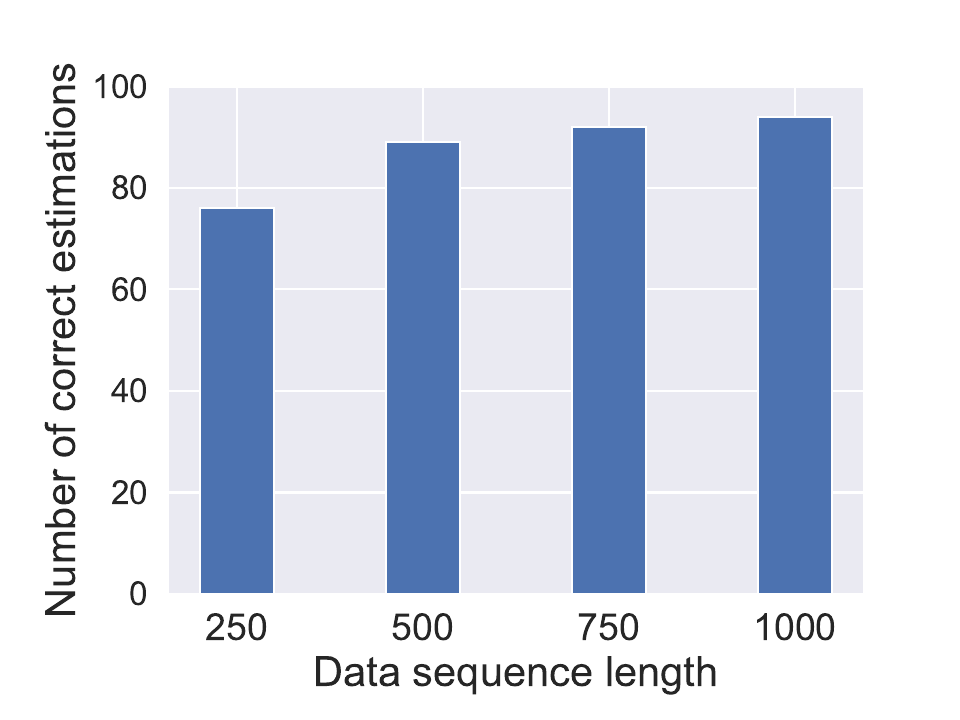}
    		\caption{$F_D$} 
                \label{sub:hist}
    \end{subfigure}\hspace{\fill}
    \caption{Sample time series with $L_o=2$ dynamical layers (a) and bar diagram of the absolute frequency of detection, $F_D$, as a function of the series length (b).}
\end{figure}

The computer experiment consists of the following steps:
\begin{itemize}
    \item[i)] Generate $R=100$ independent realisations of the time series model with $L_o=2$ described in Section \ref{ssNL}. Each signal consists of $T=1,000$ data points. A sample signal is displayed in Figure \ref{sub:signal}.
    \item[ii)] For each independent realisation, extract four subsequences containing the first 250, 500, 750 and 1,000 data points, respectively.
    \item[iii)] Generate an initial condition for each realisation, $\hat{\Lambda}_0^{(r)}$, $r=1, ..., R$. Apply the SA-EM algorithm to each subsequence of each realisation and obtain parameter estimates.  
    \item[iv)] For each $r=1, ..., R$ and each subsequence, compute the normalised errors for the transition matrix ($e_{\bm{M}}(r)$) and each real parameter ($e_{\alpha_i}(r)$), as well as the detection frequency $F_D$ for the integer delays. 
\end{itemize}
The  purpose of this setup is to demonstrate the effectiveness of the SA-EM algorithm, study the existence of local maxima of the likelihood and illustrate the improvement in the accuracy of the parameter estimates as the length of the observed series increases.

\begin{figure}[htb]
\captionsetup[subfigure]{justification=Centering}
    \begin{subfigure}[t]{0.32\textwidth}	
    		\includegraphics[width=\textwidth]{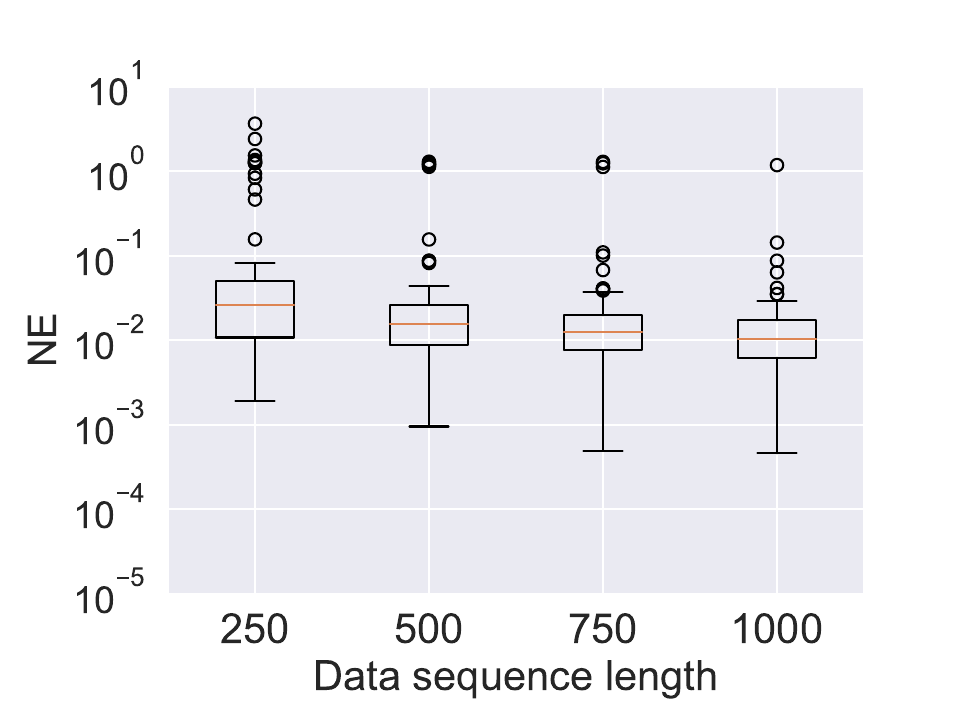}
    		\caption{Parameters $a[1:2]$.}
                \label{sub: Dis_a}
    \end{subfigure}\hspace{\fill}
    \begin{subfigure}[t]{0.32\textwidth}
    		\includegraphics[width=\linewidth]{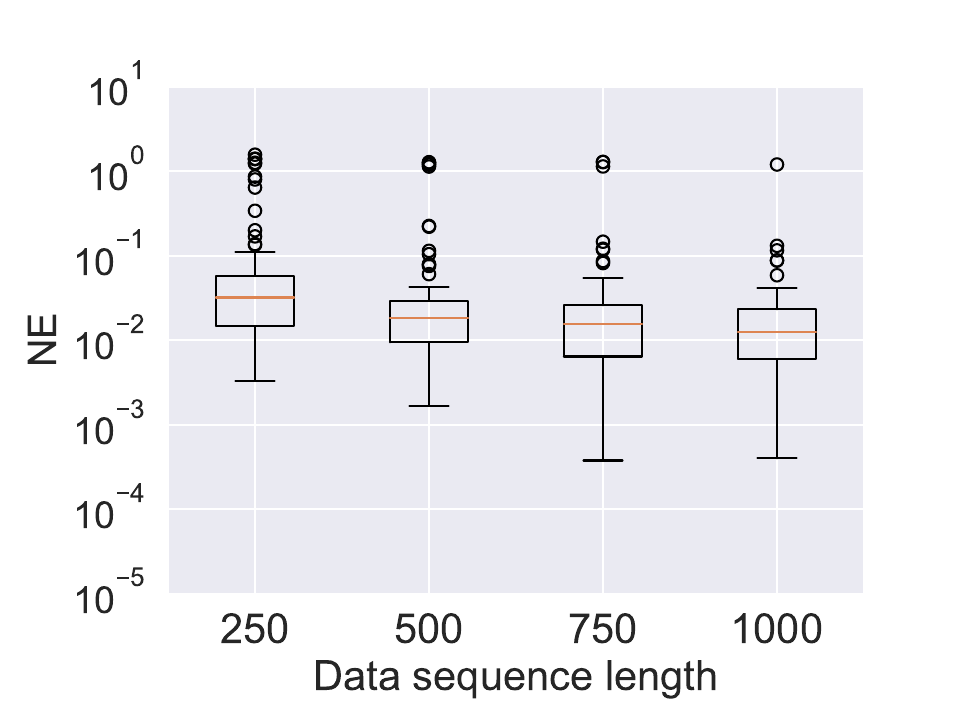}
    		\caption{Parameters $b[1:2]$.} 
                \label{sub: Dis_b}
    \end{subfigure}\hspace{\fill}
    \begin{subfigure}[t]{0.32\textwidth}
    		\includegraphics[width=\linewidth]{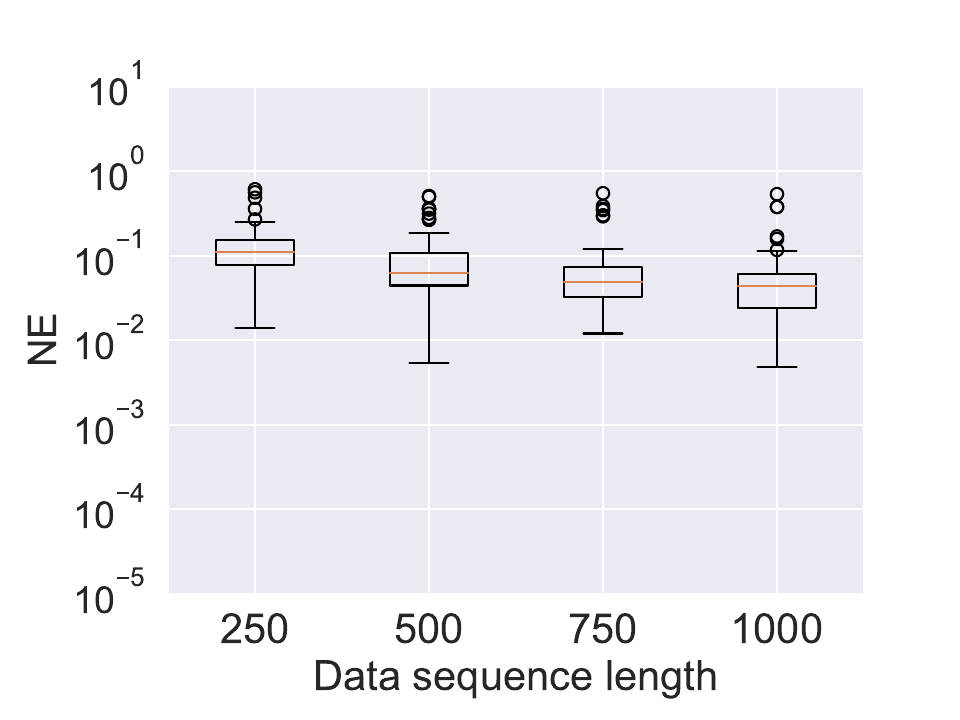}
    		\caption{Transition matrix $\bM$.}
                \label{sub: Dis_M}      
    \end{subfigure}\hspace{\fill}
    \begin{subfigure}[t]{0.32\textwidth}
    		\includegraphics[width=\linewidth]{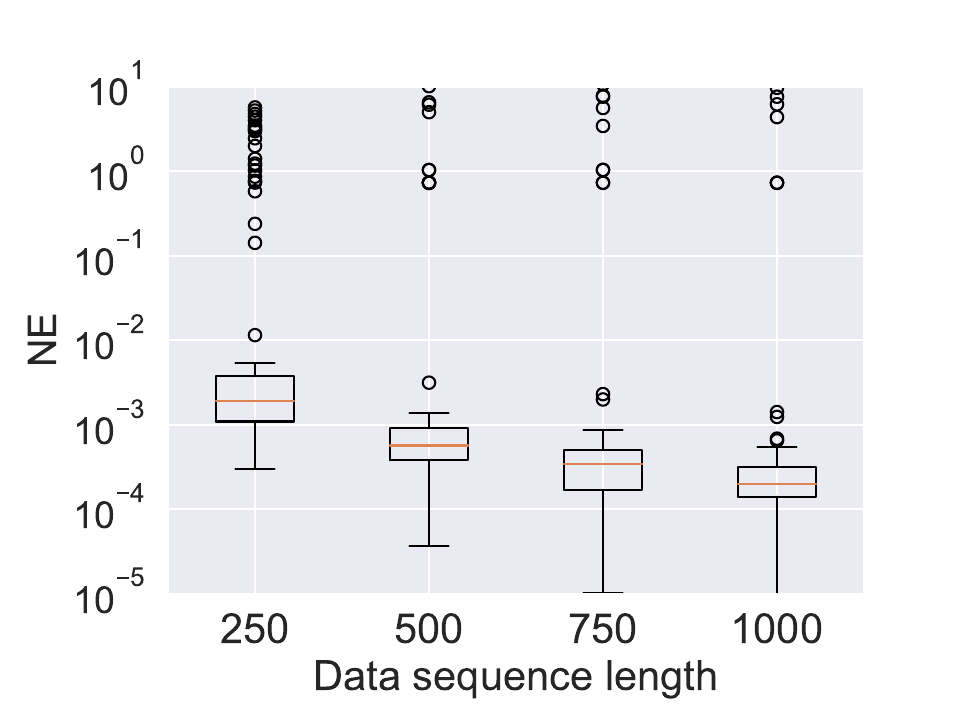}
    		\caption{Parameters $\omega[1:2]$.}
                \label{sub: Dis_w}      
    \end{subfigure}\hspace{\fill}
    \begin{subfigure}[t]{0.32\textwidth}
    		\includegraphics[width=\linewidth]{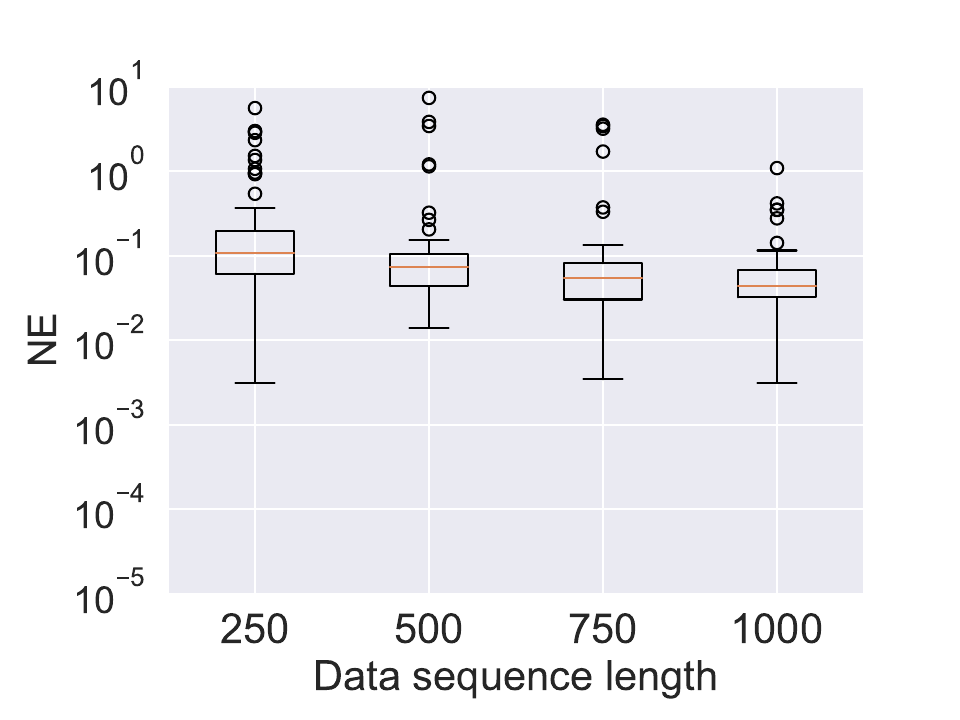}
    		\caption{Parameters $\kappa[1:2]$.}
                \label{sub: Dis_k}
    \end{subfigure}\hspace{\fill}
    \begin{subfigure}[t]{0.32\textwidth}
    		\includegraphics[width=\linewidth]{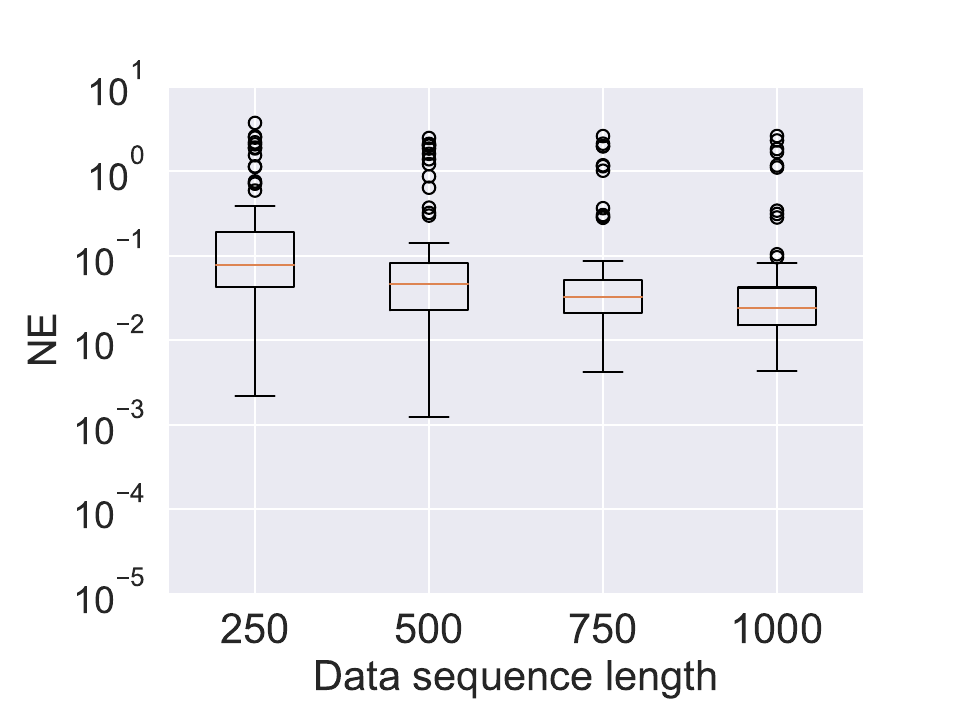}
    		\caption{Parameters $\sigma[1:2]$}
                \label{sub: Dis_s}
    \end{subfigure}
    \caption{Box-plots of the normalised errors (NE's) $e_{\bm{M}}(r)$, $e_{a[l]}(r)$, $e_{b[l]}(r)$, $e_{\kappa[l]}(r)$, $e_{\omega[l]}(r)$ and $e_{\sigma[l]}(r)$ for $r=1, \ldots, R$ and $l=1, 2$. The errors for the same parameter in the two layers, e.g., $a[1]$ and $a[2]$, are grouped in the same box plot.}
    \label{fig:Discrete}
\end{figure}

Figure \ref{sub:hist} shows a bar diagram with the absolute frequency of correct detection, $F_D$, for each data sequence length. Since there are $R=100$ simulations (for each length), the maximum value of $F_D$ is $R=100$. We see that, as the length of the data sequence increases, the value of $F_D$ improves as well. When the number of data points in the sequence is $T=1,000$ we obtain $F_D \approx 95$, i.e., $D[1]$ and $D[2]$ are both detected correctly in $\approx 95\%$ of the simulations. The delays become mismatched in the simulation runs where the SA-EM algorithm converges to a local maximum of the likelihood. 

Figure \ref{fig:Discrete} shows box plots of the normalised errors for the remaining parameters. For each parameter and each length of the data sequence, the red line indicates the median of the errors, the box extends between the .25 and .75 quantiles of the empirical distribution, the whiskers extend to the complete distribution and the circles are outliers (i.e., points located above the upper quartile by 1.5 times the interquartile range). We observe how the median error decreases when more data points are available. As with the delays, outliers are due to simulations where the SA-EM algorithm converges to a local maximum of the likelihood that differs significantly from its global maximum. These simulations indicate that, in any practical application, the SA-EM algorithm should be run with multiple initialisations (even for a single data set). One can then select the parameter estimates from the run that attained the highest log-likelihood (which is computed by the SA-EM algorithm as a by-product of the forward-backward procedure).

%
\subsection{cDN-ARMS($L$) model with non-integer delays}
\label{ssENSO_real_delays}

The assumption of integer delays, i.e., that $\tau[l]$, $l=1, \ldots, L$, are all integer multiples of the one-month time step $h=\frac{1}{12}$, is unrealistic. In this section, we assume that $D[l] = \frac{\tau[l]}{h} \in (1,+\infty)$ and construct a cDN-ARMS($L$) model of the form in Eq. \eqref{eq2} for the ENSO time series. To be specific, the SA-EM algorithm is implemented with the model
\beqa
l_n &\sim& P(l_n|l_{n-1}) = M_{l_{n-1},l_n}, \nn\\
x_n &=& x_{n-1} + h\left[ 
    b[l_n] \cos\left(
        2\pi \omega[l] h(n-1)
    \right) - a[l_n] \tanh\left(
        \kappa[l_n] \tilde{x}_{n-D[l_n]}) 
    \right)
\right] \nn \\
&& + \sqrt{h} \sigma[l_n] u_n, \quad n \ge 0,
\label{eq:ENSO-2}
\eeqa
where $\tilde{x}_{n-D[l_n]}$ is computed by linear interpolation as shown in Eq. \eqref{eqLinInterp}. The transition matrix $\bm{M}$ and the parameters $\bm{\alpha}$ are defined in the same way as in Section \ref{ssENSO_integer_delays}, and $x_n \sim \mN(0,1)$ for all $n<0$.

\subsubsection{Generation of synthetic time series} \label{sssGenData}

The interpolated signal $\tilde x_{n-D[l]}$ in \eqref{eq:ENSO-2} is an approximation that we impose to incorporate a real delay into a discrete-time model. In order to put this approximation to a test, we generate time series data using stochastic DDEs of the form in Eq. \eqref{eq:ghil08} that we integrate with an Euler-Maruyama scheme on a finer grid, namely, with a time step of the form $h=\frac{1}{12m}$, where $m$ is a positive integer. 

In particular, we let again $L_o=2$ and generate an {\em auxiliary} data sequence from the model
\beqa
y_i &=& y_{i-1} + \frac{1}{12m}\left[ 
    b[l_i] \cos\left(
        2\pi \omega[l_i] \frac{12(i-1)}{m}
    \right) - a[l_i] \tanh\left(
        \kappa[l_i] x_{n-\mD[l_i]}) 
    \right)
\right] \nn \\
&& + \sqrt{\frac{1}{12m}} \sigma[l_i] u_i, \quad i \ge 0,
\nn
\eeqa
where the $\mD[l]$'s are positive integer delays, $\{u_i\}$ is a standard Gaussian i.i.d. sequence of noise variables, $y_i \sim \mN(0,1)$ for all $i<0$, $l_i \sim P(l_i|l_{i-1})=M_{l_{i-1},l_i}$ when $i$ is an integer multiple of $m$ and $l_i=l_{i-1}$ when $i$ is {\em not} an integer multiple of $m$ (i.e., the index $l_i$ can only change every $m$ time steps).

The actual data set used for the computer simulations is then obtained by subsampling $\{y_i\}$ by a factor $m$, namely,
\beq
x_n = y_{nm}, \quad \text{for} \quad n = 0, 1, \ldots
\nn
\eeq
If $m=2$, the delays $\mD[l]$, which are integers in the discrete-time scale of $\{y_i\}$, become rational in the discrete-time scale of $\{x_n\}$, with possible values of the form $D[l] \in \left\{ r, r \pm \frac{1}{2} \right\}$, where $r \in \mbZ^+$. For general $m \in \mbZ^+$, the delays in the subsampled time scale of the series $\{ x_n \}$ are of the form $D[l] \in \left\{ r, r + \frac{1}{m}, \ldots, r+\frac{m-1}{m} \right\}$, with $r\in\mbZ^+$. In this way, we generate a data sequence $\{x_n\}$ that depends on non-integer delays and does not rely on interpolation or any other approximation based on the observed data.

\subsubsection{Parameter estimation}

We conduct a set of computer experiments similar to those in Section \ref{sssParamEst} but using data sequences generated by the procedure in Section \ref{sssGenData} to account for non-integer delays. The estimation errors for $\bm{M}$ and the parameters in vector $\alpha$ are computed in the same way as in Section \ref{sssParamEst}. The estimates of the delays for these experiments are real numbers, hence we compute normalised errors of the form $e_{D[l]}(r) = |D[l]-\hat D[l]^{(r)}|/|D[l]|$, where $D[l]$ is the true value of the delay and $\hat D[l]^{(r)}$ is the estimate in the $r$-th simulation run. 

The procedure for the computer experiments is the same as in Section \ref{sssParamEst}, with $R=100$ independently generated data sequences of length $T=1,000$, each of them split to obtain subsequences of length 250, 500, 750 and 1,000. The values of the true parameters $\bm{\alpha}$ and transition matrix $\bm{M}$ used to generate the data are the same as in the model with $L_o=2$ in Section \ref{ssNL}. The auxiliary sequence $\{y_i\}$ is generated with a time step $\frac{1}{12m}$, with $m=2$. The true non-integer delays are $D[1]=3.5$ and $D[2]=9.5$.

\begin{figure}[htb]
\captionsetup[subfigure]{justification=Centering}
    \begin{subfigure}[t]{0.32\textwidth}	
    		\includegraphics[width=\textwidth]{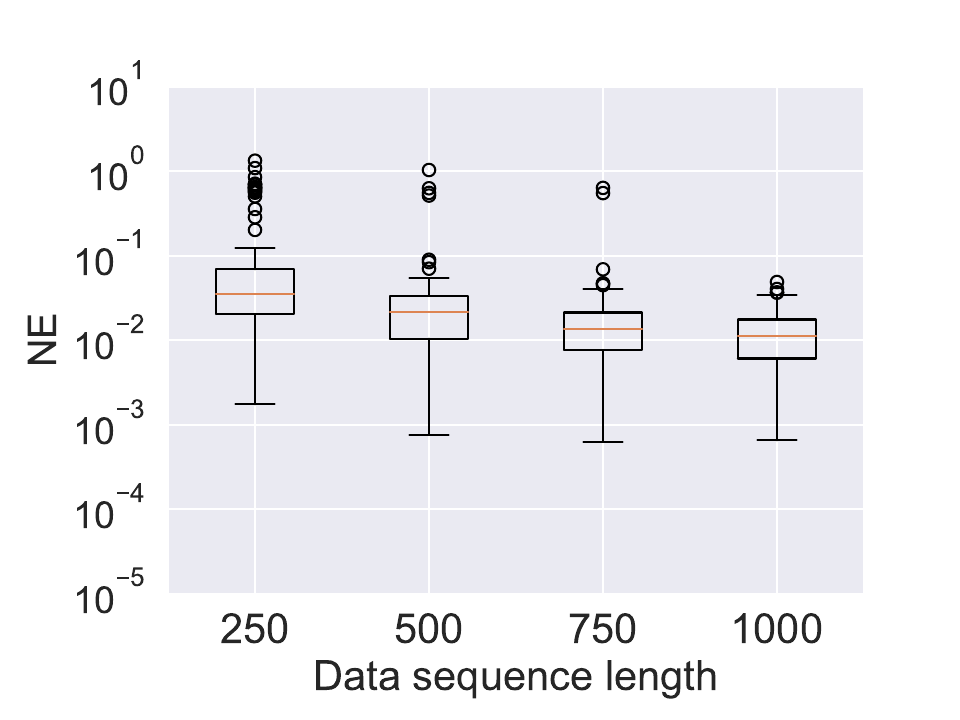}
    		\caption{Parameters $a[1:2]$.} \label{fAfrac}
    \end{subfigure}\hspace{\fill}
    \begin{subfigure}[t]{0.32\textwidth}
    		\includegraphics[width=\linewidth]{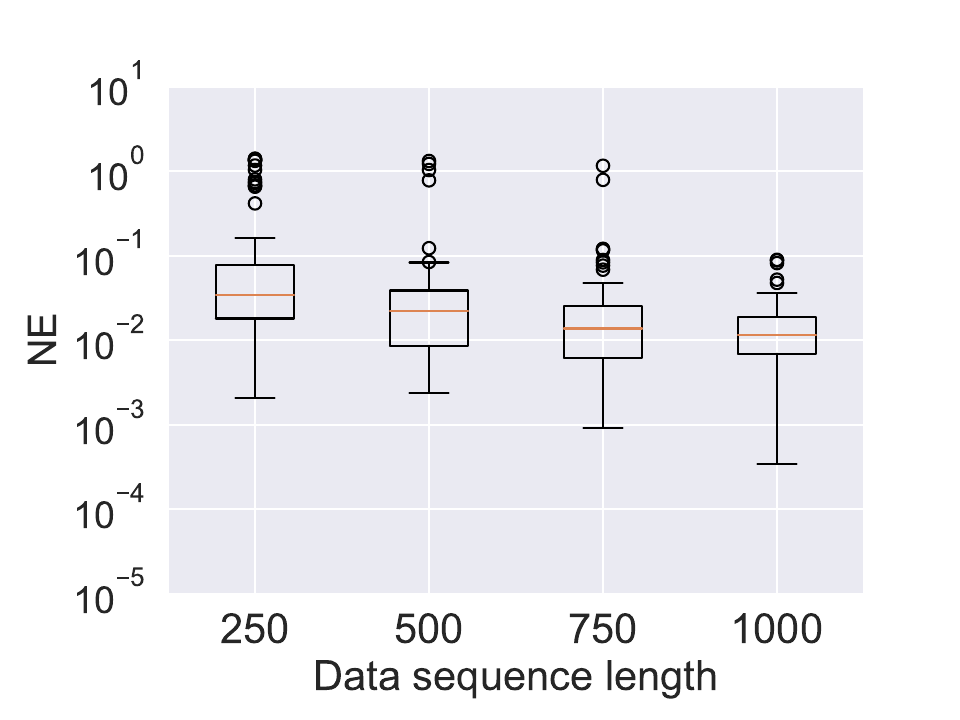}
    		\caption{Parameters $b[1:2]$.} 
    \end{subfigure}\hspace{\fill}
    \begin{subfigure}[t]{0.32\textwidth}
    		\includegraphics[width=\linewidth]{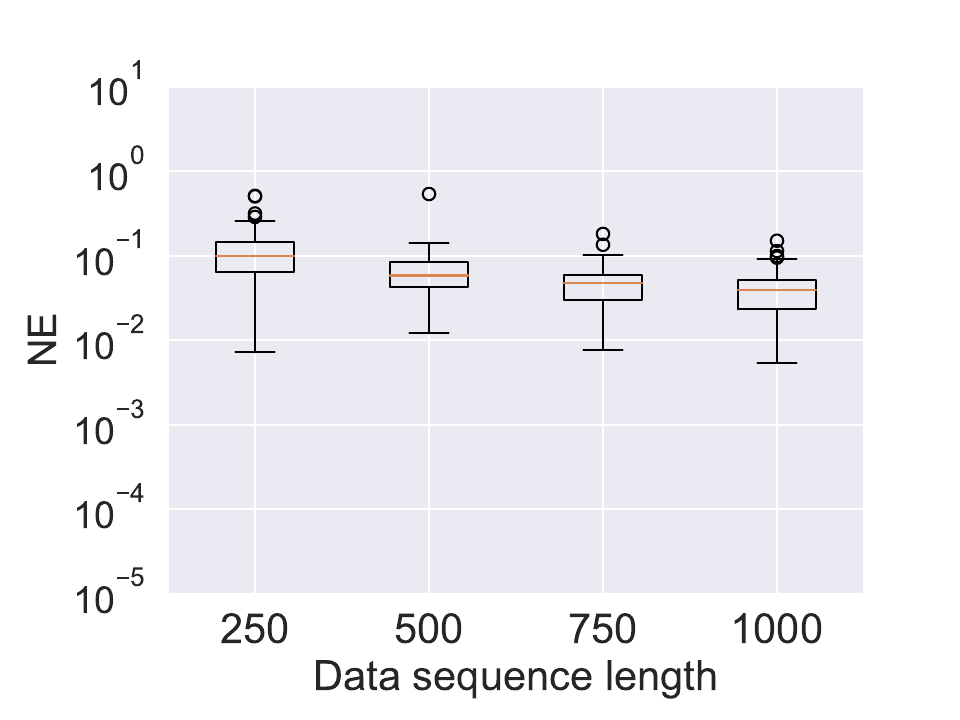}
    		\caption{Matrix $\bm{M}$.}
    \end{subfigure}\hspace{\fill}
    \begin{subfigure}[t]{0.32\textwidth}
    		\includegraphics[width=\linewidth]{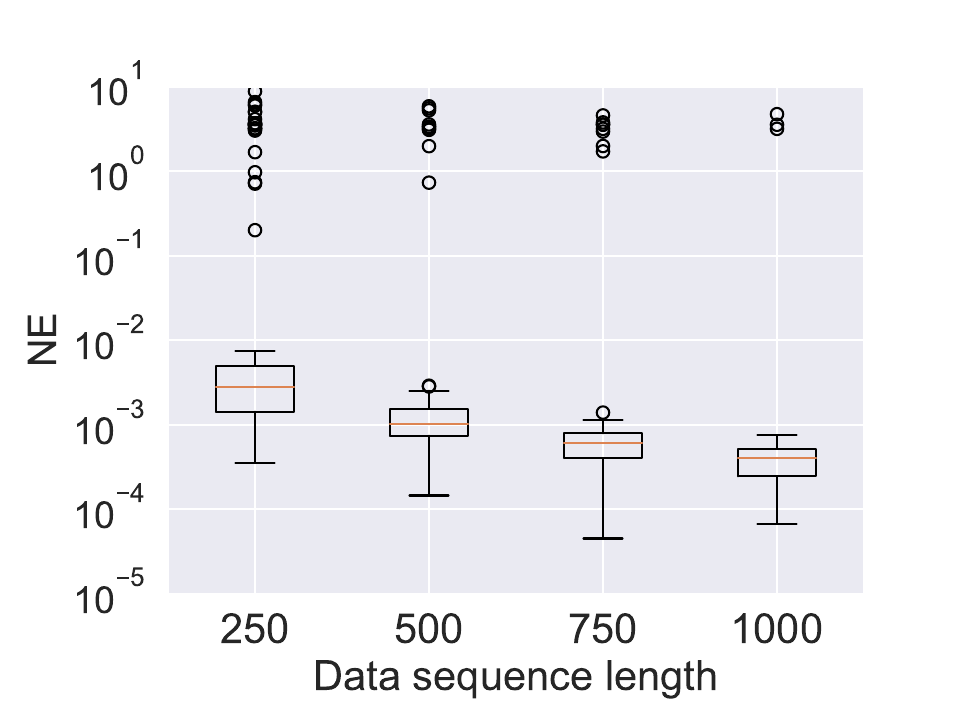}
    		\caption{Parameters $\omega[1:2]$.}
    \end{subfigure}\hspace{\fill}
    \begin{subfigure}[t]{0.32\textwidth}
    		\includegraphics[width=\linewidth]{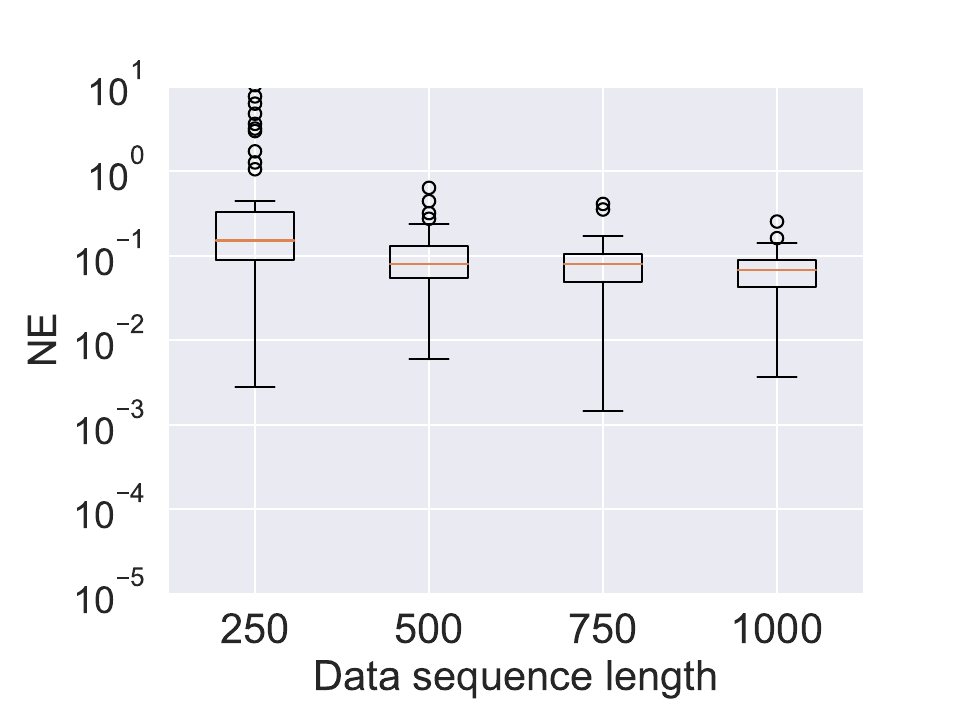}
    		\caption{Parameters $\kappa[1:2]$.}
    \end{subfigure}\hspace{\fill}
    \begin{subfigure}[t]{0.32\textwidth}
    		\includegraphics[width=\linewidth]{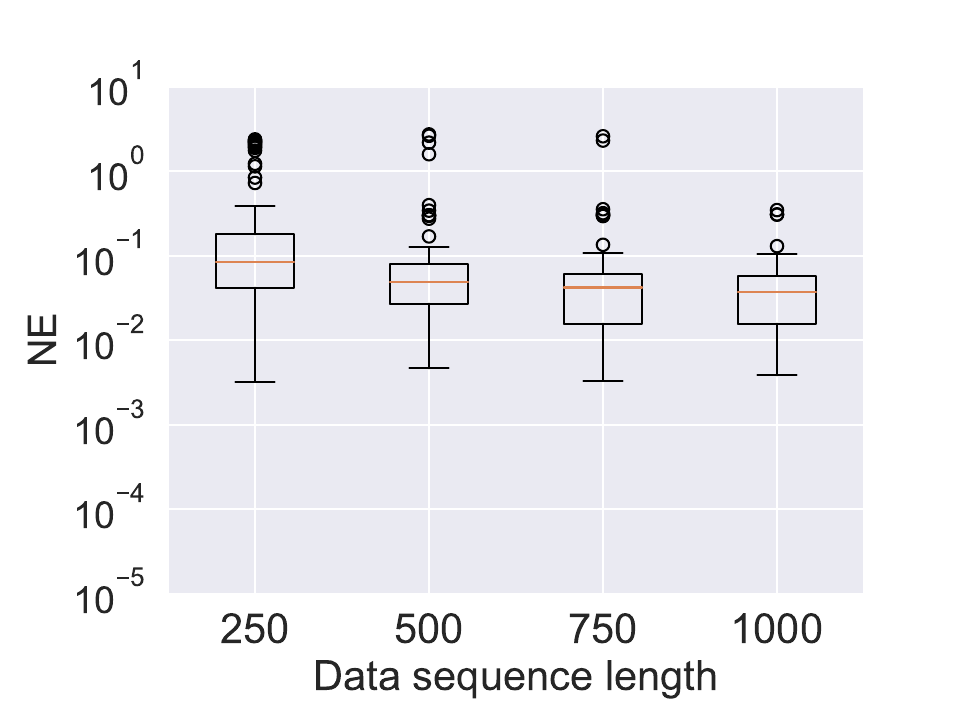}
    		\caption{Parameters $\sigma[1:2]$}
    \end{subfigure}
    \begin{subfigure}[t]{0.32\textwidth}
    		\includegraphics[width=\linewidth]{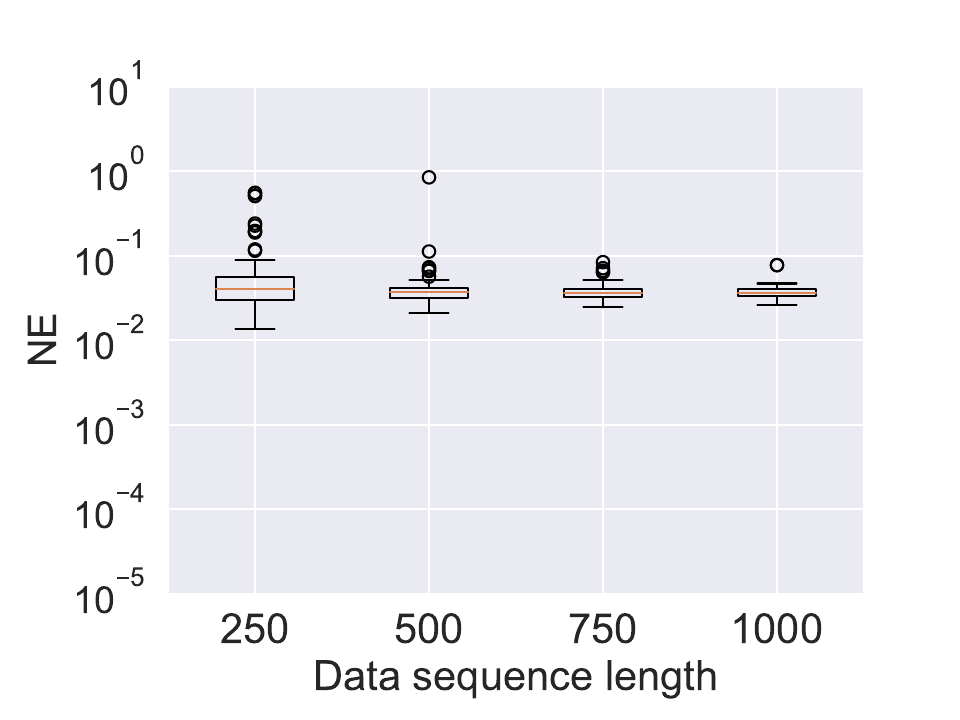}
    		\caption{Delays $D[1:2]$}
                \label{subfig:delays}
    \end{subfigure}
    \caption{Box plots of the normalised errors (NE's) $e_{\bm{M}}(r)$, $e_{a[l]}(r)$, $e_{b[l]}(r)$, $e_{\kappa[l]}(r)$, $e_{\omega[l]}(r)$, $e_{\sigma[l]}(r)$ and $e_{D[l]}(r)$ for $r=1, \ldots, R$ and $l=1, 2$.}
    \label{fig:enso}
\end{figure}

Figure \ref{fig:enso} shows the box plots of the normalised errors for all parameters. We note that the errors are grouped per parameter through the two layers, e.g., the box plot in Figure \ref{fAfrac} corresponds to the population of errors $\{ e_{a[1]}(r), e_{a[2]}(r) :  r=1, \ldots, R\}$. Similar to Figure \ref{fig:Discrete} we observe that the median errors decrease consistently as the length of the data sequence increases, except for the real delays $D[1:2]$. In this case we see that the normalised errors are small (close to $10^{-2}$) and their variance reduces with increasing data length, but the median error remains approximately constant. This numerical result indicates that the estimators of $D[1]$ and $D[2]$ present a bias due to the mismatch between model \eqref{eq:ENSO-2}, used by the SA-EM algorithm, and the model of Section \ref{sssGenData} used to generate the data. 

\newpage
\section{Model fitting with real ENSO data} \label{sRealData}

%
\subsection{Data and models}

\begin{figure}
\centerline{
	\includegraphics[width=0.45\linewidth]{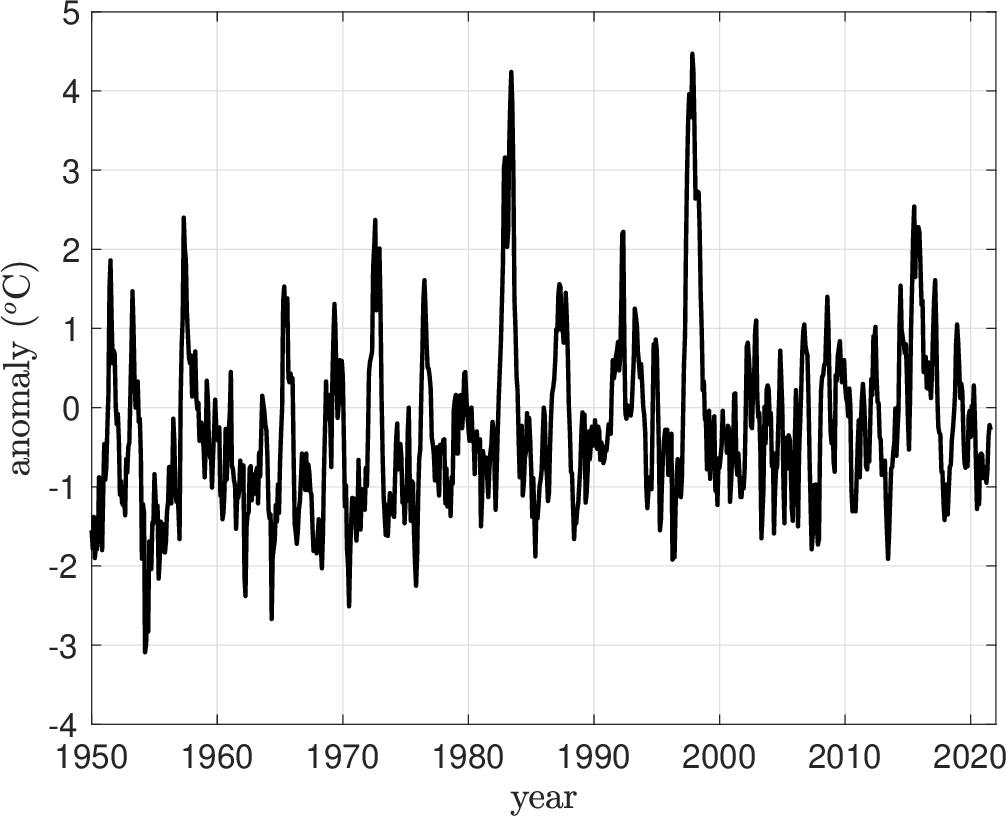}
	\includegraphics[width=0.5\linewidth]{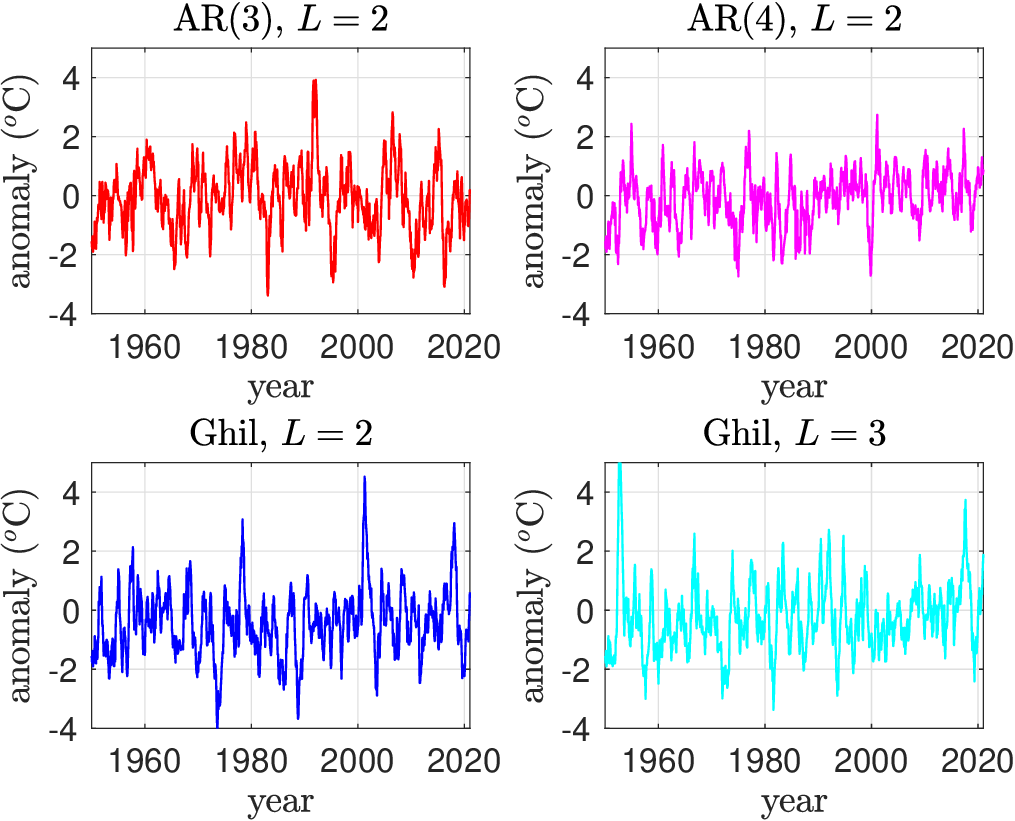}
}
\caption{\textit{Left:} ENSO zone 1+2 SST anomalies between January 1950 and August 2021 (monthly measurements). Data publicly available at \url{https://climatedataguide.ucar.edu/climate-data}. \textit{Right:} sample signals simulated using four models fitted using the real data set. \textit{Ghil, $L=2$} and \textit{Ghil, $L=3$} are the cDN-ARMS($L$) models with 2 and 3 layers, respectively, where each layer corresponds to an instance of Eq. \eqref{eq:ghil08} with different parameters. \textit{AR(3), $L=2$} and \textit{AR(4), $L=2$} are the linear AR models with $L=2$ layers.}
\label{fENSOreal}
\end{figure}

After validating the performance of the SA-EM inference algorithm with synthetic data in Section \ref{sExamples}, we now tackle the fitting of cDN-ARMS($L$) models using real ENSO data. The data set is a time series of monthly SST anomalies consisting of $T=860$ observations, starting in January 1950 and up to August 2021, shown in Figure \ref{fENSOreal} (left). This yields a sequence of $T=860$ observations.

We have applied the EM-SA algorithm on this data set in order to fit different models, namely:
\begin{itemize}
\item cDN-ARMS($L$) models with $L=2, 3, 4$,
\item two Markov-switching models with $L=2$ layers where each layer is a linear AR system; in one model each layer is AR(3) and in the second model each layer is AR(4).  
\end{itemize}
All four models are fitted using the SA-EM algorithm. For the cDN-ARMS($L$) models the algorithm is applied in the same way as in Section \ref{ssENSO_real_delays}. For the models with linear AR layers, the algorithm simplifies considerably as $\Lambda_c = \emptyset$, which implies that there is no need for the ARS optimisation scheme. The overall estimation procedure becomes very similar to the one in \cite{Franke12} (except that the models in \cite{Franke12} are of order 1).

Figure \ref{fENSOreal} (right) shows one sample series generated with with four models: two cDN-ARMS($L$) models based on Ghil et al.'s Eq. \eqref{eq:ghil08} and two Markov-switching linear AR models. We skip the cDN-ARMS($4$) model, which attained poor estimates (due to insufficient data).

%
\subsection{Parameter estimates}

The estimated transition matrices for the cDN-ARMS($2$) and cDN-ARMS($3$) models are 
\beq
\hat\bM = \left[
	\begin{array}{cc}
	0.855 & 0.145\\
	0.274 & 0.726\\
	\end{array}
\right]
\quad \text{and} \quad
\hat\bM = \left[
	\begin{array}{ccc}
	0.122 & 0.878 & 0.000\\
	0.001 & 0.898 & 0.101\\
        0.270 & 0.000 & 0.730\\
	\end{array}
\right],
\nn
\eeq
respectively. The remaining estimated parameters for these models are
\beq
\begin{array}{ll}
\hat a[1:2] = (43.953, 7.373)^\top, &\hat\kappa[1:2] = (0.050, 0.186)^\top,  \\
\hat\omega[1:2] = (0.004, 1.116)^\top, &\hat b[1:2] =(-1.550, 2.898)^\top,\\
\hat \sigma[1:2] = (1.161,1.859)^\top, &\hat D[1:2] = (2.386, 7.301)^\top\\
\end{array}
\nn
\eeq
for cDN-ARMS(2) and
\beq
\begin{array}{lll}
\hat a[1:3] = (15.555,5.372,2.982)^\top, &\hat\kappa[1:3] = (0.251,0.438,1000.000)^\top, \\
\hat\omega[1:3] = (0.670,0.003,0.121)^\top, &\hat b[1:3] =(6.830, -1.840, 2.471)^\top,\\
\hat \sigma[1:3] = (0.531,1.191,1.726)^\top, &\hat D[1:3] = (2.000,2.153,13.844)^\top
\end{array}
\nn
\eeq
for cDN-ARMS(3). We remark that, in both cases, the estimated delays have non-negligible fractional parts. This result justifies the need for modelling the delays as real random variables.



%
\subsection{Model assessment}

Table \ref{table:like-enso} shows a comparison of the four models in terms of their log-likelihood $\ell_T(L)$ and their penalised log-likelihood $\tilde \ell_T(L)$, computed as described in Section \ref{ssEstimateL}. We observe that the cDN-ARMS($L$) models attain a higher log-likelihood than the Markov-switching linear AR($L$) models and increasing the number of parameters to fit (by increasing $L$ or the order of the AR models) yields a higher likelihood. When looking at the penalised log-likelihood, however, the lower order models yield the best results and the Markov-switching linear AR(3) model with $L=2$ layers attains the highest value.  

\begin{table}[!ht]
    \captionsetup{width=1\textwidth}
    \centering
    \begin{tabular}{|l||l|l|}
    \hline
        ~ &$\ell_T(L)$ &$\tilde\ell_T(L)$\\ \hline\hline
        $Ghil,L=2$ &$-446.4$ & $\bm{-499.9}$ \\ \hline
        $Ghil,L=3$ &$\bm{-430.8}$ & -521.2 \\ \hline
        $AR(3),L=2$ & $-463.2$ & $\bm{-486.0}$ \\ \hline
        $AR(4),L=2$ & $\bm{-460.8}$ & $-507.7$ \\ \hline
    \end{tabular}
    \caption{Log-likelihoods and penalised log-likelihoods of the cDN-ARMS($L$) (Ghil) models with $L =2,3,4$ and the 2-layer linear AR models, when fitted to the real data}
    \label{table:like-enso}
\end{table}

\begin{figure}[htb]
\captionsetup[subfigure]{justification=Centering}
    \begin{subfigure}[t]{0.5\textwidth}	
    		\includegraphics[width=\textwidth]{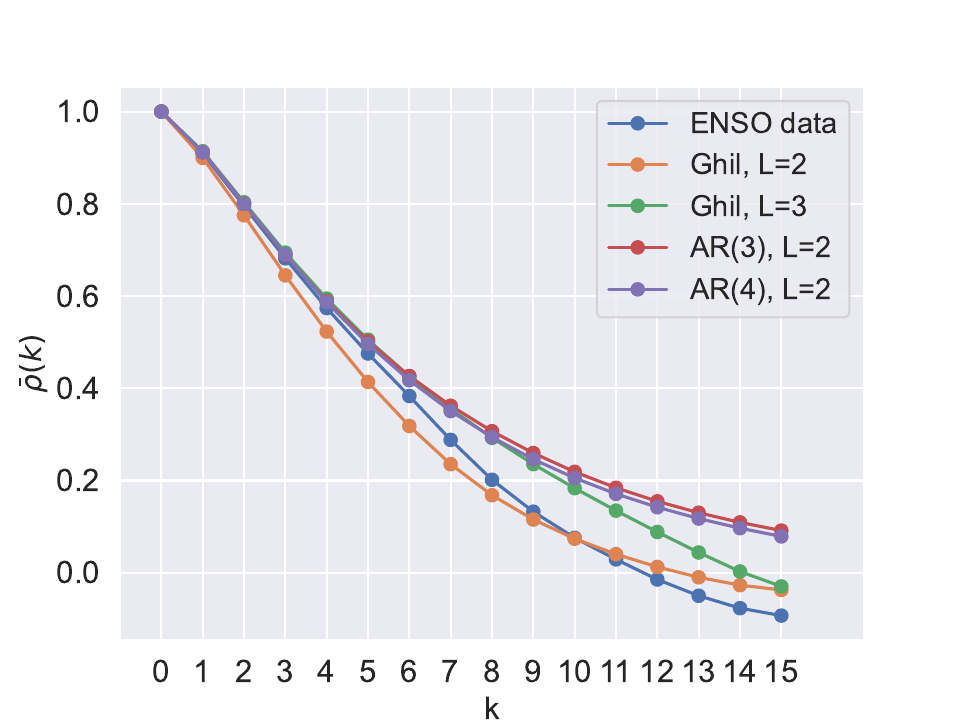}
    		\caption{Autocorrelation until $k=12$ of the real data, Ghil and AR(3) models} 
                \label{auto-plot}
    \end{subfigure}\hspace{\fill}
    \begin{subfigure}[t]{0.5\textwidth}
    		\includegraphics[width=\linewidth]{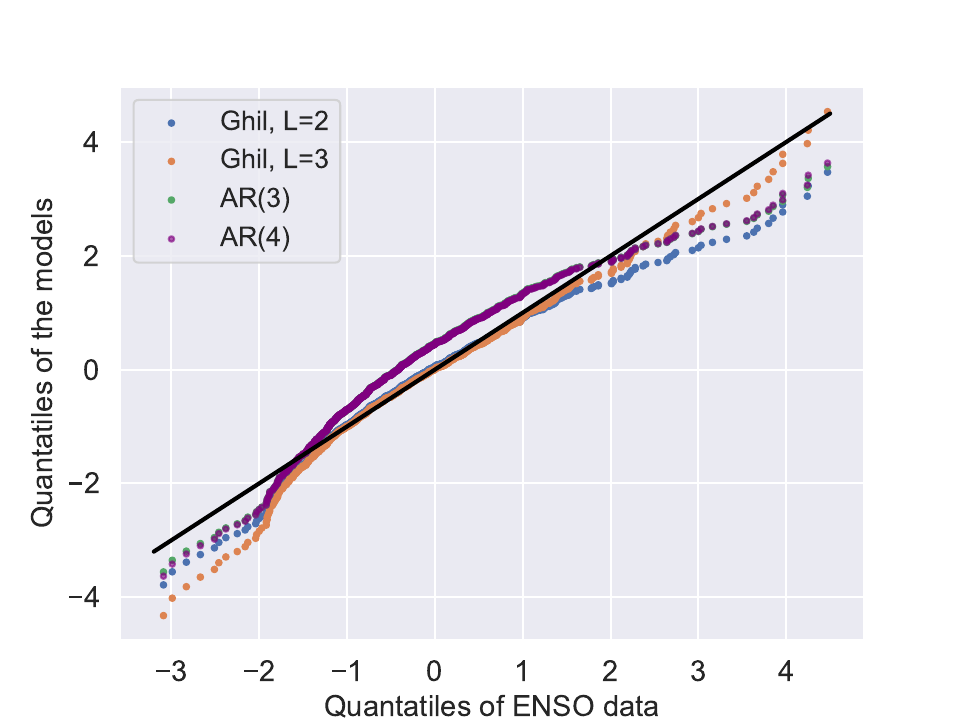}
    		\caption{QQ-plot of the AR(3) and GHIL models.} 
                \label{qq-plot}
    \end{subfigure}\hspace{\fill}
    \caption{\textit{Right:} autocorrelation functions for the real ENSO data and the fitted models from 0 to 15 months. \textit{Left:} QQ-plots of the fitted models. Models are labeled in the same way as in Table \ref{table:like-enso}.}
    \label{fCompare}
\end{figure}

Other comparisons yield a different view, however. Figure \ref{fCompare} (left) shows the empirical autocorrelation function computed using the ENSO data (blue colour) and the empirical autocorrelations obtained for the same four models in Table \ref{table:like-enso}. These curves are computed by averaging the empirical autocorrelations of 2,000 independently-generated series for each model. It is seen that the cDN-ARMS($2$) model (Ghil, $L=2$) yields the autocorrelation which is closest to the one obtained from the real data. Also, the two cDN-ARMS($L$) models capture the negative autocorrelation of the ENSO data (after month 12) although they both underestimate it. The Markov-switching models with linear AR layers fail to capture the true autocorrelation after 5-6 months. 

Finally, Figure \ref{fCompare} (right) shows the quantile-to-quantile (QQ) plots for the four models compared to the ENSO data. Here, the ENSO quantiles are obtained from the empirical distribution of the real data, while the quantiles for the models are computed by averaging over 2,000 independent simulations of each model. We observe that the Markov-switching models with linear AR layers fail to capture both the central quantiles of the distribution of the ENSO data and its tails. The two cDN-ARMS($L$) models yield a very good approximation of the central quantiles (roughly from -2 to 2) and the cDN-ARMS(3) model also yields a good approximation of the right tail of the distribution (for SST anomalies $>2$). The representation of the left tail is poorer, however. We note that positive anomalies (known as `El Ni\~no') in the ENSO data of Figure \ref{fENSOreal} (left) are stronger than the negative ones (`La Ni\~na'), which may explain why the cDN-ARMS(3) model attains a better fit of the right tail of the distribution.

%
\section{Conclusions}
\label{sConclusions}

We have introduced a class of nonlinear autoregressive Markov-switching time series models where each dynamical layer (or sub-model) may have a different, possibly non-integer delay. This class includes a broad collection of systems that result from the discretisation of stochastic DDEs where the characteristic delays are not a priori known. Such models are common in Geophysics.

The proposed family of models admits an asymptotic-regime analysis similar to the classical results of \cite{Yao00} for first-order Markov-switching systems. In particular, we have explicitly described sufficient conditions for the random sequences generated by the proposed models to converge to a unique invariant distribution. We have also introduced numerical methods, based on a space-alternating EM procedure, to detect the number of dynamical layers in the model and to compute ML estimators of any unknown parameters, including the multiple, possibly non-integer delays. The performance of these inference methods has been tested on nonlinear autoregressive Markov-switching models that combine two or three dynamical layers, each one of them originating from a DDE typically used to represent  anomalies in the sea surface temperature of (certain regions of) the Pacific Ocean due to the ENSO phenomenon. Real-world ENSO data are recorded as a monthly series, yet the delays that characterise the phenomenon are not known a priori and there is no physical reason for them to be integer multiples of a month. Our computer simulations show that it is possible to detect the number of layers and estimate the parameters of models with at least two or three dynamical layers using relatively short series of observations. The cDN-ARMS($L$) models fitted using real ENSO data and the proposed space-alternating EM algorithm can capture the autocorrelation features of the model up to 15 months and its central and right-tail quantiles. The delays estimated from the real data are, indeed, non-integer.

%
\section*{Declarations}

\subsection*{Availability of data and materials}

The datasets analysed during the current study are available in the Climate Data Guide repository of the US National Center for Atmospheric Research (NCAR). URL: \url{https://climatedataguide.ucar.edu/climate-data}

\subsection*{Conflicts of interest}

The authors declare that there are no conflicts of interest related to this research.  

\subsection*{Funding}

This work has been partially supported by the Office of Naval Research (awards N00014-22-1-2647 and N00014-19-1-2226) and {\em Agencia Estatal de Investigaci\'on} of Spain (project PID2021-125159NB-I00 TYCHE).

\subsection*{Author's contributions}

Jos\'e A. Mart\'inez-Ord\'o\~nez has contributed to the design of the work, the analysis of data, the creation of new software, and the draft and revision of the manuscript.

Javier L\'opez-Santiago has contributed to the acquisition, analysis and interpretation of data, and the revision of the manuscript.

Joaqu\'in M\'iguez has contributed to the conception and design of the work, the analysis  of data, and the draft and revision of the manuscript.

%
\begin{appendices}

\section{Accelerated random search algorithm} \label{apARS}

Let $f: \mR \subseteq \mathbb{R}^d \rightarrow [0,\infty)$ be a real objective function and consider the optimisation problem $\hat{\bm{\beta}} \in \arg \max_{\bm{\beta} \in \mR} f(\bm{\beta})$. The accelerated random search algorithm of \cite{Appel03} (see also \cite{Marinho07} for some extensions) is an iterative method for global optimisation that performs a Monte Carlo search on a sequence of balls of varying radius. The algorithm can be outlined as follows:

\begin{itemize}
\item {\em Initialisation:} choose a minimum radius $r_{\min}>0$, a maximum radius $r_{\max}>r_{\min}$, a contraction factor $c>1$ and an (arbitrary) initial solution $\bm{\beta}_0$. Set $r_1 = r_{max}$.

\item {\em Iteration:} denote the solution at $(n-1)$-th iteration as $\bm{\beta}_{n-1}$ and let $B_n:= \left\{ \bm{\beta} \in \mR: \norm{ \bm{\beta} - \bm{\beta}_{n-1} } < r_n \right\}$. To compute a new solution $\bm{\beta}_n$, take the following steps:
\begin{enumerate}
    \item Draw $\tilde{\bm{\beta}}$ from the uniform probability distribution on $B_n$.
    \item If $f(\tilde{\bm{\beta}}) > f(\bm{\beta}_{n-1})$ then set $\bm{\beta}_n = \tilde{\bm{\beta}}$ and $r_{n+1} = r_{\max}$.\\
    Otherwise, set $\bm{\beta}_n = \bm{\beta}_{n-1}$ and $r_{n+1} = \frac{r_{n}}{c}$.
    \item If $r_{n+1}<r_{\min}$, then $r_{n+1}=r_{\max}$
\end{enumerate}
\end{itemize}
The algorithm can be iterated a fixed number of times or stopped when $\bm{\beta}_n = \bm{\beta}_{n-1} = \ldots = \bm{\beta}_{n-r}$ for a prescribed, sufficiently large $r>0$.

\end{appendices}

\bibliographystyle{plain} 
\bibliography{bibliografia}

\end{document}